\documentclass[article,reqno]{amsart}
\usepackage{amsmath,amssymb}
\usepackage{latexsym}
\usepackage{epsfig}
\usepackage{cite}
\usepackage{booktabs}
\usepackage{color}
\usepackage{graphicx}
\usepackage{tikz}
\usepackage{pgfplots}
\newcommand{\tens}{\pmb}

\numberwithin{equation}{section}

\title{Exact quantisation of the relativistic Hopfield model}
\author{F. Belgiorno$^{1,2}$ \and S.L. Cacciatori$^{3,4}$ \and F. Dalla Piazza$^{5}$ \and M. Doronzo$^{3}$}
\address{\noindent $^1$Dipartimento di Matematica, Politecnico di Milano, Piazza Leonardo 32, IT-20133 Milano, Italy\endgraf
$^2$INdAM-GNFM \endgraf
$^3$Department of Science and High Technology, Universit\`a dell'Insubria, Via Valleggio 11, IT-22100 Como, Italy\endgraf
$^4$INFN sezione di Milano, via Celoria 16, IT-20133 Milano, Italy\endgraf
$^5$Universit\`a ``La Sapienza'', Dipartimento di Matematica, Piazzale A. Moro 2, I-00185, Roma, Italy}
\begin{document}
\maketitle

\begin{abstract}
We investigate the quantisation in the Heisenberg representation of a relativistically covariant version of the Hopfield model for dielectric media, 
which entails the interaction of the quantum electromagnetic field with the matter dipole fields. The matter fields are represented by a mesoscopic polarization field. 
A full quantisation of the model is provided in a covariant gauge, with the aim of maintaining explicit relativistic covariance. 
Breaking of the Lorentz invariance due to the intrinsic presence in the model of a preferred reference frame is also taken into account. 
Relativistic covariance forces us to deal with 
the unphysical (scalar and longitudinal) components of the fields, furthermore it introduces, in a more tricky form, the well-known dipole ghost of 
standard QED in a covariant gauge. In order to correctly dispose of this contribution, we implement a generalized 
Lautrup trick. Furthermore, causality and the relation of the model with the Wightman axioms are also discussed. 
\end{abstract}

\section{Introduction}

Interaction of the quantum electromagnetic field with quantum matter represents a huge and longstanding field of investigation, which 
can involve both a microscopic description of the fields and a more phenomenological approach in which some 
microscopic interactions are described by means of effective fields, which in turn can be quantised and coupled with QED through 
effective interaction vertices. An example of this kind of approach is provided by models describing interactions of the 
electromagnetic field with dielectric media. Interest in this framework has been recently risen up, due to the attempt to 
reproduce quantum emission by a black hole in the lab by means of analogous systems, i.e. systems displaying the same kinematics which is at the root of the 
Hawking effect \cite{hawking-cmp,unruh-seminal,barcelo,philbin}.   
A possible implementation of the analogue Hawking effect as a quantum effect in dielectric media has been presented and analysed in \cite{PRL2010-H}, \cite{PRL2011-Answer}, \cite{NJP2011} and further developed in \cite{petev}.  Improvement of  
such theoretical analysis required to look for a more fundamental model, able to maintain the main aspects of the phenomenology of the system, and still providing a good mathematical model.\\
A simple and economic model describing the interaction of the electromagnetic field with a homogeneous and isotropic medium is the Hopfield model \cite{Hopfield1958}. With economic we mean that this model allows to reproduce the
observed spectrum of the electromagnetic field into a class of transparent dielectric media (the Sellmeier dispersion relations) without entering the details of a complete physical description of the materials themselves: they are described
as a lattice (or a field) of independent oscillators all having a given (one or more) characteristic frequency $\omega_0$. Thus, the model does not record any detail about the matter apart from the characteristic frequency of the dipoles.
Nevertheless, the Hopfield model in its original form is not able to accomplish the preposed targets. Indeed, the analogue Hawking radiation and other perturbative and non perturbative phenomena occur in presence of inhomogeneities propagating through 
a homogeneous background (see e.g. \cite{philbin}). These phenomena require in general to be able to move from the lab frame to another inertial one, such as the frame comoving with a perturbation or to a carrier signal. 
For this reason we have developed a relativistic covariant version of the Hopfield model in \cite{PhysicaScripta}. In \cite{EPJD2014} we have considered the quantisation in the lab frame in a simple fixed gauge, in order to study  
photon production originated by time-dependent perturbations. 
The covariant and gauge invariant quantisation has been considered in \cite{PhysicaScripta}, where in the construction of states   the linear coupling between the electromagnetic field and the polarisation field was treated perturbatively. The Hawking effect 
as a non perturbative effect  has been also considered in a simplified model in \cite{PRD2015}.\\
As the action of the full covariant Hopfield model is quadratic in all fields, we expect that the theory may be quantisable  in an exact way in the Heisenberg representation. This is the aim of the present paper. It is worth noting that this is 
a relevant task for several reasons. First, the fact that the perturbative quantisation seems to be well-defined and causal, does not ensure that the same properties hold true for the full theory, and difficulties may arise in the exact theory in the Heisenberg representation,
which remain hidden in the formally well working perturbative construction (but which would appear after adding further perturbative terms). For example, let us consider two Klein-Gordon scalar fields linearly interacting, having the Lagrangian
density
\begin{eqnarray}
L=\frac 12 \Box \phi_1-\frac {m_1^2}2\phi_1 +\frac 12 \Box \phi_2-\frac {m_2^2}2 \phi_2-\lambda^2 \phi_1 \phi_2.
\end{eqnarray}
The equations of motion in the Fourier dual space are
\begin{eqnarray}
\begin{pmatrix}
-\tens{k}^2+m_1^2 & \lambda^2 \\ \lambda^2 & -\tens{k}^2 +m_2^2
\end{pmatrix}
\begin{pmatrix}
\tilde \phi_1 \\ \tilde \phi_2
\end{pmatrix}
=
\begin{pmatrix}
0 \\ 0
\end{pmatrix},
\end{eqnarray}
where we use the symbol $\tens{k}=(k^0, \vec k)$ for indicating a spacetime vector, whereas $\vec k$ indicates a spatial vector. In order to provide nontrivial solutions, the matrix has to have vanishing determinant, so that one
gets the dispersion relation
\begin{eqnarray}
(\tens k^2)^2-(m_1^2+m_2^2)\tens k^2 +m_1^2m_2^2-\lambda^4=0.
\end{eqnarray}
Since the discriminant is always non negative, this equation has two real solutions in $\tens k^2$. However, we see that one of these solutions will be negative if the condition
\begin{eqnarray}
\lambda^2\leq m_1m_2
\end{eqnarray}
is not satisfied. This means nothing else that, if $\lambda^2>m_1m_2$,  then the exact theory after a linear recombination of the fields, decouples in a pair of fields, one bradionic and the other tachyonic. For example, if one of the
original masses is zero, say $m_1=m$, $m_2=0$, then the theory is affected by the presence of tachyon states, irrespectively of the smallness of the coupling constant w.r.t. $m$. It is not evident that this fault of the model could be easily recovered in the 
perturbative formulation. Presence of the tachyonic modes shows an instability in the theory, due to the fact that the  perturbation is not bounded from below unless it is small enough w.r.t. both the masses.\\
The covariant version of the Hopfield model is found in a similar situation: we are considering a massless field, the photon, coupled to an oscillator field with a linear coupling which is, a priori, non necessarily small, and, surely not at all small with respect to the 
zero photon mass. Still, we shall find that the covariant Hopfield model behaves differently and the theory remains well defined for all values of the coupling constant. \\
A second reason justifying the relevance of the exact quantisation is the following. The dipole field is defined as a field of identical oscillators which, in the frame where the matter is at rest, satisfy the harmonic equation
\begin{eqnarray}
\frac {\partial^2 \vec P}{\partial t^2} +\omega_0^2 \vec P=\vec 0.
\end{eqnarray}
Now, this equation introduces two conceptual difficulties. The first one is that it is not a priori relativistic: if we would interpret the dipoles as linearly proportional to the displacement of the centre of mass of the electron w.r.t. to the one of
the nuclei, for example, then, since the equation of motion is linear, nothing prevents both the product $\omega_0^2 \vec P$ and then the speed of the moving clouds of electrons, to be arbitrarily large, also compared to the speed of light. This could
give rise to problems with the causality properties of the theory. On the other hand, the fact that in the model the microscopical details of the structure of matter are irrelevant, at least as far as low energy scales are involved, 
suggests that it is not affected by such microscopical aspects and it might 
keep satisfying relativistic causality.
For this reason it is crucial to check the consistency of the theory by verifying
that causality is not violated. At this point enters the second conceptual complication. Indeed, causality can be easily proved as a byproduct of the compatibility of the equal time canonical commutation relations (ETCCR) and  relativistic covariance.
Since the dipole equations are defined in a specific frame, the one where the dielectric is at rest, in any other frame the velocity $\tens v$ specifying the frame will enter the equations of motion. This is nothing but the fact that Lorentz 
invariance is broken by the presence of the dielectric medium, which selects a preferred inertial reference frame. This does not implies that also the covariance is broken, but it makes the question more subtle, since it requires
for relativistic covariance to be realised correctly in the construction of the representation of the algebra of fields. Again, we shall see that this is the case for the relativistic Hopfield model, so that it preserves both covariance and causality.\\
Beyond the above mentioned questions, further difficulties arise in the full construction, as the control of gauge invariance, the presence of the dipole ghost and the right choice of the test functions space necessary to define the fields as operator
valued distributions.\\ 
The plan of the paper is as follows:  in sec. \ref{sec:classical} we consider the classical relativistic Hopfield model with a single polarisation field and the set of its solutions in the Fourier dual space. This allows us to identify
a suitable space of test functions. The presence of a dipole ghost also emerges. In standard electromagnetic field quantisation, as, for example, in QED, the presence of the dipole ghost can be 
avoided by working in the Feynman gauge. In our case the analogue of the Feynman gauge does not exist. In order to take the dipole ghost under control, in sec. \ref{sec:Lautrup} we adapt to our model a strategy due to Lautrup \cite{Lautrup1967}, 
which is generalised in a Lorentz covariant form. In sec. \ref{sec:Fourier modes}, we introduce the quantum fields and use the conserved scalar product naturally associated with the Hamiltonian structure, in order
to determine the Fourier modes and the corresponding commutation relation. In particular, we find the explicit expression of the Hamiltonian operator. 
In sec. \ref{sec:covariance and causality}
we discuss how the covariance can be realised in the construction of the Fock representation of the quantum algebra and, in particular, how it ensures causality. Indeed, the presence of the dielectric medium breaks the Lorentz invariance
and this does not allow to represent the full covariant theory on a single Fock space. We need a family of Fock spaces 
which are related to each other by unitary maps. 
On each of such Fock spaces we get the realisation of the theory in  
any inertial frame in which the dielectric medium has velocity $\tens{v}$. Each Fock space is invariant under the action of the little group of $\tens v$, the subgroup of the Poincar\'e group leaving $\tens v$ invariant.
In sec. \ref{sec: propagator} we compute the two point function;  being the theory Gaussian, this is sufficient for reconstructing the whole theory. Finally, in sec. \ref{sec: conlusions} we resume all results, by rewriting them in a more compact 
way, using a more suitable language. There, we also discuss possible perspectives.\\
We mention that some of the results presented here have been anticipated by us in \cite{Quantum-fipsi}, in a simplified form for a more elementary model in which the electromagnetic field and the polarisation field are replaced by scalar fields, so that
several technical complications, such as gauge invariance and the presence of constraints, are absent. However, no proves are presented there, only the ideas, and all technical details and mathematical proofs will be found herein. 

\

Finally, we add some comments on notations: we shall use the symbols $\tens x, \tens k$ or $\tens V$ for the space-time vectors having components $x^\mu, k^\mu$ and $V^\mu$, whereas their spatial component will be indicated with
$\vec x, \vec k$, $\vec V$ and so on. We shall use $k^2$ for the scalar $\tens k^2=\tens k \cdot \tens k$. In our units the speed of light is $c=1$.

\section{The relativistic Hopfield model and its solutions}\label{sec:classical}
Let us consider the relativistic Hopfield model with a single polarisation field with resonance frequency $\omega_0$, as presented in \cite{PhysicaScripta}. The Lagrangian density is
\begin{align}
\mathcal L=&-\frac 1{16\pi} F_{\mu\nu} F^{\mu\nu} -\frac 1{2\chi\omega_0^2}(v^\rho \partial_\rho P_\mu)(v^\sigma \partial_\sigma P^\mu)+\frac 1{2\chi} P_\mu P^\mu-\frac {g}{2} (v_\mu P_\nu-v_\nu P_\mu) F^{\mu\nu}+B{\partial_\mu A^\mu}+ \frac \xi2 B^2,
\end{align}
so that the equation of motion are
\begin{align}
\frac 1{4\pi} (\Box \eta_{\mu\nu}- \partial_\mu \partial_\mu)A^\mu +g ( \eta_{\mu\nu}v^\rho \partial_\rho -v_\mu \partial_\nu)P^\nu-\partial_\mu B &=0, \\
g (\eta_{\mu\nu}v^\rho \partial_\rho - v_\nu \partial_\mu )A^\nu-\frac 1{\chi \omega_0^2} (\omega_0^2+(v^\rho \partial_\rho)^2) P^\mu &=0, \\
\partial_\mu A^\mu+\xi B &=0,
\end{align}
together with the defining constraint $v^\mu P_\mu=0$.
After defining the Fourier transforms of the fields
\begin{align}
\tilde{\tens A}(\tens k)&=\int_{\mathbb R^4}d^4x e^{i\tens k \cdot \tens x} \tens A(\tens k), \\
\tilde{\tens P}(\tens k)&=\int_{\mathbb R^4}d^4x e^{i\tens k \cdot \tens x} \tens P(\tens k), \\
\tilde B(\tens k)&=\int_{\mathbb R^4}d^4x e^{i\tens k \cdot \tens x} B(\tens k),
\end{align}
the system of partial differential equations transform into the algebraic system
\begin{eqnarray}
\mathcal M V\equiv 
\begin{pmatrix}
\frac 1{4\pi} (k^2\mathbb I-\tens k\otimes \tens k) & ig (\omega \mathbb I -\tens v\otimes \tens k ) & -i\tens k \\
-i g (\omega \mathbb I -\tens k\otimes \tens v ) & \frac 1{\chi_0} \left( \frac {\omega^2}{\omega_0^2}-1 \right)\mathbb I & 0 \\
i\tens k & 0 & -\xi 
\end{pmatrix}
\begin{pmatrix}
\tilde {\tens A} \\ \tilde {\tens P} \\ \tilde B
\end{pmatrix}
=
\begin{pmatrix}
\tens 0 \\ \tens 0 \\ 0
\end{pmatrix},
\end{eqnarray}
where $\omega:=k^\mu v_\mu$. We need to solve this system of equations in a distributional sense. The determinant of the matrix $\mathcal M$ defining the system is 
\begin{eqnarray}
\det \mathcal M=-\frac {(k^2)^2}{\chi^4} \left( \frac {\omega^2}{\omega_0^2}-1 \right)^4\left[\frac {k^2}{4\pi} -\frac {g^2\chi \omega_0^2 \omega^2}{\omega^2-\omega_0^2} \right]^2
\left[\frac {1}{4\pi} -\frac {g^2\chi \omega_0^2 }{\omega^2-\omega_0^2} \right].
\end{eqnarray}
Thus, the solutions will have support in the union of four sets. The first set is defined by $k^2=0$, the same spectrum of free photons in vacuum. The associated modes of the photon field do not couple with matter, and
are expected to appear as nonphysical states. The second set is defined by $\omega^2=\omega_0^2$: since $\omega_0$ is the resonance frequency, these solutions correspond to modes where the dipoles of the polarisation field oscillate freely. We shall see
that these modes are projected out by the physical constraint $\tens v \cdot \tens P=0$. The next branch is defined by
\begin{eqnarray}
\frac {1}{4\pi} -\frac {g^2\chi \omega_0^2 }{\omega^2-\omega_0^2}=0.
\end{eqnarray}
This mode looks like a resonance at a shifted frequency 
\begin{eqnarray}
\bar \omega=\omega_0 \sqrt {1+4\pi g^2\chi}.
\end{eqnarray}
and it is not particularly significative, as it is shown to correspond to an uncoupled contribution to the polarisation field. Finally, there are the modes satisfying the equation
\begin{eqnarray}
\frac {k^2}{4\pi} -\frac {g^2\chi_0 \omega_0^2 \omega^2}{\omega^2-\omega_0^2}=0.
\end{eqnarray}
These modes reproduce exactly the Sellmeier relation with just one resonance, so that they represent the physical modes we are interested to.\\
In order to find the solutions we can make use of the identity
\begin{eqnarray}
x\delta^{(n+1)}(x)=-(n+1)\delta^{(n)}(x),
\end{eqnarray}
where $\delta^{(n)}$ is the $n$-th derivative of the Dirac delta function $\delta=\delta^{(0)}$. Assuming that generically $\tens k$ and $\tens v$ are linearly independent and defining the four-vectors $\tens e_1$ and $\tens e_2$, satisfying 
$\tens e_i\cdot \tens k=0$, $\tens e_i\cdot \tens v=0$ and such that $\tens k, \tens v, \tens e_1, \tens e_2$ is a basis for $\mathbb R^4$, we find for the general solution of the linear system:
\begin{eqnarray}
&& \tilde {\tens A}(\tens k)=(\omega^2-\omega_0^2) a_0(\tens k) \tens v \delta (k^2)+\omega \left[ \omega^2-\omega_0^2-\frac \xi{4\pi} (\omega^2-\bar\omega^2) \right] a_0(\tens k) \tens k \delta^{(1)}(k^2)+a_3(\tens k) \tens k \delta (k^2)\cr
&&\phantom{ \pmb {\mathcal A}(\pmb k)=}-\frac {4\pi}{k^2}(\omega \tens k -k^2\tens v) g b_3(\tens k) \delta (\omega^2-\bar \omega^2)+\sum_{i=1}^2 \delta\!\left(\frac {k^2}{4\pi} -\frac {g^2\chi \omega_0^2 
\omega^2}{\omega^2-\omega_0^2} \right) a_i(\tens k) \tens e_i, \label{A}\\
&& \tilde {\tens P}(\tens k)=ig {\chi \omega_0^2} (\omega \tens v -\tens k) a_0(\tens k) \delta(k^2)
+i(\omega \tens v - \tens k) b_3(\tens k) \delta (\omega^2-\bar\omega^2)\cr
&& \phantom{\pmb {\mathcal P}(\pmb k)=}+ig \sum_{i=1}^2 \frac {\chi \omega_0^2}{\omega^2-\omega_0^2} \omega  \delta\!\left(\frac {k^2}{4\pi} -\frac {g^2\chi \omega_0^2 \omega^2}{\omega^2-\omega_0^2} \right)a_i(\tens k) \tens e_i,\\
&& \tilde B(\tens k)=i  \frac \omega{4\pi} (\omega^2-\bar\omega^2) a_0(\tens k) \delta (k^2).
\end{eqnarray}
There is also an additional mode for $\tilde {\tens P}$ proportional to $\delta(\omega-\omega_0)$, which is excluded by the condition $\tens v\cdot \tilde{\tens P}=0$. 
At this point, we can be a little bit more precise on specifying the solutions space. The Fourier transforms are supported on the sets
\begin{align}
\Sigma_0:=&\{\tens k \in \mathbb R^{1,3}| k^2=0 \},\\
\Sigma_1:=&\{\tens k \in \mathbb R^{1,3}| \omega^2=\bar\omega^2\},\\
\Sigma_2:=&\{\tens k \in \mathbb R^{1,3}| \frac {k^2}{4\pi} -\frac {g^2\chi \omega_0^2 \omega^2}{\omega^2-\omega_0^2}=0\}.
\end{align}
At first sight, we can take the coefficients in the set $\mathcal S(\mathbb R^4)$ of rapidly decreasing smooth functions. However, a problem arises because these supports have a non vanishing intersection. We shall see in section
\ref{sec: propagator} that $\Sigma_2$ intersect $\Sigma_1$ and $\Sigma_0$ in two isolated points, which give no rise to any singularity. Instead, $\Sigma_0\cap \Sigma_1$ consists in a two-dimensional ellipsoid. 
We require for the Fourier coefficient to vanish on this intersection locus at the appropriate degree. The presence of the $\delta^{(1)}(k^2)$ suggests that the order should be two. We shall confirm this in section \ref{sec: propagator}. \\
Let us discuss shortly the solutions. The presence of the
$\delta^{(1)}(k^2)$ distribution is related to the fact that the determinant of $\mathcal M$ vanishes at second order in $k^2$. This also happens for the Sellmeier modes, but one can check that no $\delta^{(1)}$ contributions are present in that case.
The $\delta^{(1)}(k^2)$ term is well known already for the free electromagnetic field, and, indeed, it is not a surprise for it to appear in the sector $k^2=0$. This is the dipole ghost, responsible for the infrared divergent terms for the
propagators in a covariant gauge different from the Feynman gauge \cite{nakanishi,greiner}. One practical way to manage it in 
standard QED consists in choosing the Feynman gauge, $\xi=4\pi$, where the $\delta^{(1)}$ term disappears and one is dealt with the standard Fock construction
where only the spurious harmonic gauge modes are involved. In our case the dipole ghost term is proportional to
\begin{eqnarray}
1-\frac \xi{4\pi} \frac {\omega^2-\bar \omega^2}{\omega^2-\omega_0^2},
\end{eqnarray}
which cannot be eliminated by any choice of the gauge parameter $\xi$. So, we need a different strategy for managing the dipole ghost. The key observation is to note that the dipole ghost term is proportional to $\tens k$, this meaning
that it is a pure gauge mode, corresponding to a vanishing field strength and thus can be hopefully made harmless in some way. This is essentially the strategy adopted by Lautrup in \cite{Lautrup1967} for constructing the Fock representation
for the electromagnetic field in any covariant gauge. We shall now adapt Lautrup's method to our case.

\section{The generalised Lautrup strategy}\label{sec:Lautrup}
A strategy for getting rid of the dipole ghost term is as follows. First we note that the $\delta^{(1)}$ term is proportional to $\tens k$, which means that in the configurations space it is of the form $\partial_\mu f(\vec x,t)$ for some $f$. 
In an unconstrained theory, this would be a pure gauge configuration (it does not contribute to the electromagnetic field). However, because of the introduction of the auxiliary field $B$, the whole gauge symmetry is broken and only 
harmonic gauge transformation, that are transformations of the form $A_\mu\rightarrow A_\mu+\partial_\mu \Lambda$ with $\Lambda$ harmonic ($\Box \Lambda=0$), are allowed as symmetry transformation. Thus, there is no way to eliminate 
the undesired term by a symmetry gauge transformation in the constrained theory. Nevertheless, we can always separate this pure gauge term from the rest by decomposing the field $\tens A$ as follows:
\begin{eqnarray}
\tens A(\tens x)=\tens A^{F}(\tens x)+ \tens {grad} \sigma(\tens x),
\end{eqnarray}
for some scalar function $\sigma$, in such a way that the $\delta^{(1)}$ contribution is contained in $\tens {grad} \sigma$. The suffix $F$ stays for Feynman, for analogy with the usual case. 
To this aim, we need to look at the very origin of the $\delta^{(1)}$ solution. By working in the momentum space, we have $\tilde {\tens A}=\tilde {\tens A}^F-i\tilde \sigma \tens k$, where $\tilde \sigma$ is the Fourier transform of $\sigma$, and the equations
of motion take the form
\begin{align}
\frac 1{4\pi} k^2 \tilde{\tens A}^F -\left[\frac 1{4\pi} k^2 \tilde \sigma -\left( \frac \xi{4\pi}-1\right)\tilde B \right] i \tens k+i g \left[ \omega \tilde {\tens P}-(\tens k\cdot \tilde {\tens P})\tens v \right]&=\tens 0,\\
\frac 1{\chi} \left( \frac {\omega^2}{\omega_0^2}-1 \right) \tilde {\tens P} -ig \left( \omega \tilde {\tens A}^F -(\tens v\cdot \tilde {\tens A}^F) \tens k \right) &=\tens 0, \\
\tens k\cdot \tilde {\tens A}&=-i\xi \tilde B.
\end{align}
Being $\omega^2\ne \omega_0^2$, as $\omega^2= \omega_0^2$ is associated with a mode $\tilde {\tens P}\propto \tens v$ that is forbidden by the physical constraint $\tens v \cdot \tens P=0$, we can write
\begin{eqnarray}
\tilde {\tens P}=i g \frac {\chi \omega_0^2}{\omega^2-\omega_0^2} \left(  \omega \tilde {\tens A}^F -(\tens v\cdot \tilde {\tens A}^F) \tens k \right),
\end{eqnarray}
and, then,
\begin{eqnarray}
&& \!\!\!\!\!\!\!\!\! \frac {k^2}{4\pi} \tilde {\tens A}^F -\left[\frac {k^2}{4\pi}  \tilde \sigma -\left( \frac \xi{4\pi}-1\right)\tilde B \right] i \tens k-\frac {g^2 \chi \omega_0^2}{\omega^2-\omega_0^2} 
\left[ \omega^2 \tilde {\tens A}^F-\omega (\tens v \cdot \tilde {\tens A}^F) \tens k- \omega (\tens k \cdot \tilde {\tens A}^F) \tens v +k^2 (\tens v \cdot \tilde {\tens A}^F) \tens v \right]=\tens 0. \cr && \label{formulazza}
\end{eqnarray}
It is now convenient to proceed by looking at solution in the different supports determined above, branch by branch. 
Let us start with the solutions having support in $k^2=0$, which is the branch containing the dipole ghost.  We assume that $\tilde \sigma$ and $\tilde B$ are any two given scalar distributions such that
\begin{eqnarray}
\left[\frac 1{4\pi} k^2 \tilde \sigma -\left( \frac \xi{4\pi}-1\right)\tilde B \right] i \tens k=: U(\tens k)i\tens k \delta(k^2)
\end{eqnarray}
so that only the $\delta$ appears. Thus, substituting the ansatz  
\begin{eqnarray}
\tilde {\tens A}^F(\tens k)=a(\tens k) \delta(k^2) \tens k+b(\tens k) \delta^{(1)}(k^2) \tens k+r(\tens k) \delta(k^2) \tens v+\sum_{i=1}^2 d_i(\tens k) \delta(k^2) \tens e_{(i)}(\tens k),
\end{eqnarray}
in (\ref{formulazza}), we get $d_i(\tens k)=0$, whereas $a(\tens k)$ and $r(\tens k)$ can be arbitrary, and the coefficient of the $\delta^{(1)}$ is
\begin{eqnarray}
b(\tens k)=-4\pi i U(\tens k)+4\pi \frac {g^2\chi \omega_0^2 \omega}{\omega^2-\omega_0^2} r(\tens k).
\end{eqnarray}
We can eliminate the $\delta^{(1)}$ from $\tilde {\tens A}^F$ by choosing $\tilde \sigma$ in such a way that the coefficient $b(\tens k)$ vanishes, which is equivalent to say that we have to cancel out all terms linear in $\tens k$ in (\ref{formulazza}). Thus, 
the desired decomposition is defined by the condition
\begin{eqnarray}
\frac 1{4\pi} k^2 \tilde \sigma -\left( \frac \xi{4\pi}-1\right)\tilde B +i \frac {g^2 \chi \omega_0^2}{\omega^2-\omega_0^2} \omega (\tens v \cdot \tilde {\tens A}^F)=0. \label{gauge-condition}
\end{eqnarray}
At this point, the equation for $\tilde {\tens A}^F$ is
\begin{eqnarray}
&& \!\!\!\!\!\!\!\!\! \frac {k^2}{4\pi} \tilde{\tens A}^F -\frac {g^2 \chi \omega_0^2}{\omega^2-\omega_0^2} 
\left[ \omega^2 \tilde {\tens A}^F- \omega (\tens k \cdot \tilde {\tens A}^F) \tens v +k^2 (\tens v \cdot \tilde {\tens A}^F) \tens v \right]=\tens 0,
\end{eqnarray}
which has general solution
\begin{align}
\tilde{\tens A}^F(\tens k)=&(\omega^2-\omega_0^2)a_0(\tens k) \delta (k^2) \tens v +\sum_{i=1}^2 \delta\!\left(\frac {k^2}{4\pi} -\frac {g^2\chi \omega_0^2 \omega^2}{\omega^2-\omega_0^2} \right) a_i(\tens k) \tens e_i+
\delta\!\left(\frac {k^2}{4\pi} -\frac {g^2\chi \omega_0^2 \omega^2}{\omega^2-\omega_0^2} \right)\tilde a_3(\tens k) \tens k \cr
&+4\pi g b_3(\tens k) \delta (\omega^2-\bar \omega^2) \tens v,  \label{quarantadue}
\end{align}
which corresponds to a sum of plane wave solutions.\\
For the polarisation vector we get
\begin{align}
\tilde {\tens P}(\tens k)=&ig {\chi \omega_0^2 } a_0(\tens k) \delta (k^2) (\omega \tens v-\tens k)+ig \frac {\chi \omega_0^2 \omega}{\omega^2-\omega_0^2} \sum_{i=1}^2 \delta\!\left(\frac {k^2}{4\pi} 
-\frac {g^2\chi \omega_0^2 \omega^2}{\omega^2-\omega_0^2} \right) a_i(\tens k) \tens e_i\cr
&+i b_3(\tens k) \delta (\omega^2-\bar \omega^2) (\omega \tens v-\tens k).
\end{align}
Notice that from
\begin{eqnarray}
&& \tilde B=\frac i{\xi} \tens k \cdot \tilde {\tens A} = \frac i{\xi} \tens k \cdot \tilde {\tens A}^F+\frac i\xi k^2 \tilde \sigma 
\end{eqnarray}
and using (\ref{gauge-condition}), we get
\begin{eqnarray}
\tilde B(\tens k)=\frac i{4\pi} (\tens k \cdot \tilde {\tens A}^F) -i \frac {g^2\chi \omega_0^2 \omega}{\omega^2-\omega_0^2} (\tens v \cdot \tilde {\tens A}^F),
\end{eqnarray}
and so
\begin{eqnarray}
\tilde B(\tens k)=(\omega^2-\bar\omega^2)i \frac \omega{4\pi} a_0(\tens k)\delta(k^2).
\end{eqnarray}
In particular, $B$ is harmonic, as expected. Finally, we have to solve the equation for $\tilde \sigma$
\begin{eqnarray}
k^2\tilde \sigma(\tens k)=(\xi-4\pi)\tilde B(\tens k) -4\pi i \omega \frac {g^2\chi \omega_0^2}{\omega^2-\omega_0^2} (\tens v\cdot \tilde {\tens A}^F), 
\end{eqnarray}
that is
\begin{align}
k^2\tilde \sigma(\tens k)=&\left(\frac \xi{4\pi} (\omega^2-\bar\omega^2)-\omega^2+\omega_0^2\right) i \omega a_0(\tens k)\delta(k^2)-ik^2 \tilde a_3(\tens k) \delta\!\left(\frac {k^2}{4\pi} 
-\frac {g^2\chi \omega_0^2 \omega^2}{\omega^2-\omega_0^2} \right) \cr
&-4\pi i g \omega b_3(\tens k)  \delta (\omega^2-\bar\omega^2). \label{chi-equation}
\end{align}
In order to find the right representation, let us simplify the expressions by setting 
\begin{eqnarray}
V(\tens k):=\left(\frac \xi{4\pi} (\omega^2-\bar \omega^2)-\omega^2+\omega_0^2\right) i \omega a_0(\tens k)
\end{eqnarray}
and look for a particular solution of the equation
\begin{eqnarray}
k^2\tilde \sigma(\tens k)=V(\tens k) \delta(k^2).
\end{eqnarray}
Since $(k^2)^2 \tilde \sigma(\tens k)=0$, if we are working in the space of tempered distributions, the solution must be of the form
\begin{eqnarray}
\tilde \sigma(\tens k)=\alpha(\tens k) \delta(k^2)+\beta(\tens k)\delta^{(1)}(k^2), 
\end{eqnarray}
for some scalar functions $\alpha$ and $\beta$.
Here is exactly where the $\delta^{(1)}$ enters the game. Note that the $\alpha$ part is harmonic, so it is not determined by the equation, which, indeed, simply gives
\begin{eqnarray}
\beta(\tens k)=-V(\tens k).
\end{eqnarray}
In the configuration space, let us give a look at the particular solution
\begin{eqnarray}
\sigma(\tens x)\equiv -\int_{\mathbb R^4} \frac {d^4k}{(2\pi)^4} V(\tens k) e^{-i\tens k\cdot \tens x} \delta^{(1)}(k^2),
\end{eqnarray}
where with the symbol $\equiv$ we mean modulo harmonic terms, which can be reabsorbed in the pure harmonic gauge terms, corresponding to the $a_3$ term in (\ref{A}). If we consider the identity
\begin{eqnarray}
\delta^{(1)}(k^2)=\frac 1{2\omega} v^\mu \frac {\partial }{\partial k^\mu} \delta (k^2)  
\end{eqnarray}
we can write
\begin{eqnarray}
\sigma(\tens x)\equiv -\frac i2 \int_{\mathbb R^4} \frac {d^4k}{(2\pi)^4} V(\tens k) \frac {\tens v \cdot \tens x}{\omega} e^{-i\tens k\cdot \tens x} \delta(k^2).
\end{eqnarray}
After inserting back the expression of $V$ we get
\begin{eqnarray}
\sigma(\tens x)\equiv \frac 12 \int_{\mathbb R^4} \frac {d^4k}{(2\pi)^4} \left(\frac \xi{4\pi} (\omega^2-\bar \omega^2)-\omega^2+\omega_0^2\right)  a_0(\tens k) (\tens v \cdot \tens x) e^{-i\tens k\cdot \tens x} \delta(k^2),
\end{eqnarray}
which is a covariant version of the function $\Lambda$ introduced by Lautrup in \cite{Lautrup1967}. The corrections due to the remaining terms in (\ref{chi-equation}) are\footnote{Notice that, 
in the limit $g\to0$, the last row disappears and the second one becomes harmonic.}
\begin{align}
\sigma(\tens x)\equiv& \frac 12 \int_{\mathbb R^4} \frac {d^4k}{(2\pi)^4} \left(\frac \xi{4\pi} (\omega^2-\bar\omega^2)-\omega^2+\omega_0^2\right)  a_0(\tens k) (\tens v \cdot \tens x) e^{-i\tens k\cdot \tens x} \delta(k^2) \cr
&-i \int_{\mathbb R^4} \frac {d^4k}{(2\pi)^4} \tilde a_3(\tens k) e^{-i\tens k\cdot \tens x} \delta\!\left(\frac {k^2}{4\pi}-\frac {g^2\chi \omega_0^2 \omega^2}{\omega^2-\omega_0^2} \right) \cr
&-4\pi i g   \int_{\mathbb R^4} \frac {d^4k}{(2\pi)^4}\frac {\omega}{k^2} b_3(\tens k) e^{-i\tens k\cdot \tens x}   \delta (\omega^2-\bar\omega^2).
\end{align}
In adding $\tens {grad} \sigma$ to $ {\tens A}^F$, the terms with $\tilde a_3$ in (\ref{quarantadue}) cancel out. Thus, we apparently loose a degree of freedom. However, $\sigma$ is defined modulo additive harmonic terms, as
we said, which can be equally included in $ {\tens A}^F$ without changing its property of excluding the dipole ghost. For this reason we prefer to drop the $\tilde a_3$ term out from $\sigma$, and substitute the corresponding term in (\ref{quarantadue}) 
with the harmonic term $a_3(\tens k) \delta(k^2)\tens k$, whereas we include, for later convenience, a pure harmonic gauge term in $\tens {grad}\sigma$, whose Fourier transform is
\begin{eqnarray}
-\frac {i}{4\omega_+} \frac {\frac \xi{4\pi} (\omega^2-\bar\omega^2)^2+\left(\omega^2-\omega_0^2\right) (\omega^2-\bar\omega^2) -
8\pi \omega_+^2 g^2 \chi \omega_0^2}{(\omega^2-\bar\omega^2)(\omega^2-\omega_0^2)}a_0(\tens k)\tens k \delta(k^2). \label{shift}
\end{eqnarray}
In conclusion, we write
\begin{eqnarray}
\tens A(\tens x)=\tens A^F(\tens x) +\tens{grad} \sigma,
\end{eqnarray}
with 
\begin{align}
\tens A^F(\tens x)=&\int_{\mathbb R^4} \frac {d^4 k}{(2\pi)^4}\left[(\omega^2-\omega_0^2) a_0(\tens k) \delta (k^2) \tens v +\sum_{i=1}^2 \delta\!\left(\frac {k^2}{4\pi} -\frac {g^2\chi \omega_0^2 \omega^2}{\omega^2-\omega_0^2} \right) a_i(\tens k) \tens e_i+
\delta\!\left( {k^2} \right) a_3(\tens k) \tens k \right. \cr
&\left.\phantom{\sum_{i=1}^2\left(\frac {k^2}{4\pi} \right)}+4\pi g b_3(\tens k) \delta (\omega^2-\bar\omega^2) \tens v\right] e^{-i\tens k\cdot \tens x}, 
\end{align}
and
\begin{align}
\tens {grad}\sigma(\tens x)=&\int_{\mathbb R^4} \frac {d^4 k}{(2\pi)^4} \left[
\frac 12 \left(  \frac \xi{4\pi}(\omega^2-\bar\omega^2)-\omega^2+\omega_0^2 \right) (\tens v-i (\tens v\cdot \tens x) \tens k) +\right. \cr
&\left. -i  \frac {\frac \xi{4\pi} (\omega^2-\bar\omega^2)^2+\left(\omega^2-\omega_0^2\right)  (\omega^2-\bar\omega^2)-
8\pi {\omega^2 g^2 \chi \omega_0^2}}{4\omega(\omega^2-\bar\omega^2)}\tens k \right] \delta(k^2) a_0(\vec k) e^{-i\tens k \cdot \tens x} \cr
&-4\pi  c   \int_{\mathbb R^4} \frac {d^4k}{(2\pi)^4}\frac {\omega}{k^2} \delta (\omega^2-\bar\omega^2)\tens k b_3(\tens k) e^{-i\tens k\cdot \tens x}.
\end{align}
In particular, we note that $\tens A^F$ contains a basis of plane waves. Let us notice that the above  redefinition of the fields introduces a singularity in $\omega^2=\bar\omega^2$, so one could be prevented in doing it. However, as
we have discussed in section \ref{sec:classical}, such a singular locus is expected to appear in any case, as we shall see, because of the non vanishing intersection of the different branches.

\section{Scalar products and Fourier modes}\label{sec:Fourier modes}
We can now write down the whole explicit solution. To this end we can eliminate the delta distributions by integrating out the zeroth component of $\tens k$. In doing this, we must consider separately each of the supports of the deltas and look at the 
corresponding spectrum. Let us look shortly at the various spectra:
\begin{enumerate}
\item The support $k^2=0$ corresponds to the two branches $k^0(\vec k)=\pm |\vec k|$. We shall use the notation $k_\pm$ to indicate the corresponding functions of $\vec k$. More in general, with the notation $f_\pm (\tens k)$ we mean that
$k^0$ is evaluated at $k^0_\pm(\vec k)$. The property $k^0_+(\vec k)=-k^0_-(\vec k)$ can be employed in order to get real solutions. 
\item The support $(\tens k \cdot \tens v)^2-(\bar\omega)^2=0$ has the two branches 
\begin{eqnarray}
k^0(\vec k)=\frac {\vec k\cdot \vec v\pm \bar \omega}{v^0}.
\end{eqnarray}
We shall reserve the name $k^0_>$ for the solution corresponding to the positive sign, which is the positive solution in the lab frame. More in general, given a function $f(\tens k)$ we shall write $f_>(\vec k):=f(k^0_>(\vec k),\vec k)$. If with $k^0_<$
we indicate the solution with the minus sign, then $k^0_<(-\vec k)=-k^0_>(\vec k)$. In particular, $\omega_>=\bar\omega$.
\item The support
\begin{eqnarray}
\frac {k^2}{4\pi} -\frac {g^2\chi \omega_0^2 \omega^2}{\omega^2-\omega_0^2}=0
\end{eqnarray}
defines four branches, two with positive $\omega$ and two with negative $\omega$. If we call $k^0_{(a)}$, $a=1,2$, the two solutions corresponding to positive values of $\omega$, then the solutions with negative $\omega$ are
$-k^0_{(a)}(-\vec k)$. Moreover, if we order $k^0_{(a)}$ so that $k^0_{(1)}> k^0_{(2)}$, then one has  $k^0_{(1)}>\omega_0> k^0_{(2)}$. See \cite{Quantum-fipsi} for these properties. Again, we shall use the notation 
$f_{(a)}(\vec k)=f(k^0_{(a)}(\vec k), \vec k)$ for any function $f(\tens k)$. Finally, we introduce the notations
\begin{eqnarray}
DR:=\frac {k^2}{4\pi} -\frac {g^2\chi \omega_0^2 \omega^2}{\omega^2-\omega_0^2}
\end{eqnarray}
and indicate its derivative w.r.t. $k^0$ by
 \begin{eqnarray}
DR'=\frac {dDR}{dk^0}=\frac {k^0}{2\pi} +2\omega v^0 \frac {g^2\chi\omega_0^4}{(\omega^2-\omega_0^2)^2}.
\end{eqnarray}
\end{enumerate}
Thus, the explicit solution of the original system can be written in the form
\begin{align}
\tens A(\tens x) =& \int_{\mathbb R^3} \frac {d^3\vec k}{(2\pi)^3} \frac {a_0(\vec k)}{2|\vec k|} \left[ (\omega_+^2-\omega_0^2)\tens v +\frac 12 \left(\frac \xi{4\pi} (\omega_+^2-\bar\omega^2)-\omega_+^2+\omega_0^2 \right) 
(\tens v-i (\tens v\cdot \tens x) \tens k_+) \right]
e^{-i\tens k_+ \cdot \tens x}\cr 
& -i \int_{\mathbb R^3} \frac {d^3 \vec k}{(2\pi)^3} \frac {a_0(\vec k)}{2|\vec k|} \frac {\frac \xi{4\pi} (\omega_+^2-\bar\omega^2)^2+\left(\omega_+^2-\omega_0^2\right) (\omega_+^2-\bar\omega^2)-
8\pi {\omega_+^2 g^2 \chi \omega_0^2}}{4\omega_+ (\omega_+^2-\bar\omega^2)}\tens k_+  e^{-i\tens k_+ \cdot \tens x} \cr
&+i\int_{\mathbb R^3} \frac {d^3\vec k}{(2\pi)^3} \frac {a_3(\vec k)}{2|\vec k|}  \tens k_+e^{-i\tens k_+ \cdot \tens x}\cr
& +4\pi g \int_{\mathbb R^3} \frac {d^3\vec k}{(2\pi)^3}  \frac {b_3(\vec k)}{2\bar\omega v^0} \left( \tens v -\frac {\bar \omega}{k_>^2} \tens k_> \right) e^{-i\tens k_> \cdot \tens x}+\sum_{a=1}^2 \sum_{i=1}^2 
\int_{\mathbb R^3} \frac {d^3\vec k}{(2\pi)^3} \frac { a^{(a)}_i(\vec k)}{DR'_{(a)}} \tens e_i^{(a)} e^{-i\tens k_{(a)} \cdot \tens x} \cr
& +\int_{\mathbb R^3} \frac {d^3\vec k}{(2\pi)^3} \frac {a^\dagger_0(\vec k)}{2|\vec k|} \left[ (\omega_+^2-\omega_0^2) \tens v +\frac 12 \left( \frac \xi{4\pi} (\omega_+^2-\bar\omega^2)-\omega_+^2+\omega_0^2 \right) (\tens v+i (\tens v\cdot \tens x) \tens k_+) \right]
e^{i\tens k_+ \cdot \tens x}  \cr
& +i \int_{\mathbb R^3} \frac {d^3 \vec k}{(2\pi)^3} \frac {a^\dagger_0(\vec k)}{2|\vec k|} \frac {\frac \xi{4\pi} (\omega_+^2-\bar\omega^2)^2+\left(\omega_+^2-\omega_0^2\right) (\omega_+^2-\bar\omega^2)-
8\pi {\omega_+^2 g^2 \chi \omega_0^2}}{ 4\omega_+(\omega_+^2-\bar\omega^2)}\tens k_+  e^{i\tens k_+ \cdot \tens x} \cr 
&-i\int_{\mathbb R^3} \frac {d^3\vec k}{(2\pi)^3} \frac {a^\dagger_3(\vec k)}{2|\vec k|}  \tens k_+e^{i\tens k_+ \cdot \tens x}\cr
& +4\pi g \int_{\mathbb R^3} \frac {d^3\vec k}{(2\pi)^3}  \frac {b^\dagger_3(\vec k)}{2\bar\omega v^0} \left( \tens v -\frac {\bar \omega}{k_>^2} \tens k_> \right) e^{i\tens k_> \cdot \tens x}+\sum_{a=1}^2 \sum_{i=1}^2 
\int_{\mathbb R^3} \frac {d^3\vec k}{(2\pi)^3} \frac { a^{(a)\dagger}_i(\vec k)}{DR'_{(a)}} \tens e_i^{(a)} e^{i\tens k_{(a)} \cdot \tens x}, \label{campo A}
\end{align}
\begin{align}
\tens P(\tens x) =& i g \chi \omega_0^2 \int_{\mathbb R^3} \frac {d^3\vec k}{(2\pi)^3} \frac {a_0(\vec k)}{2|\vec k|}  (\omega_+ \tens v - \tens k_+)e^{-i\tens k_+ \cdot \tens x}
+i \int_{\mathbb R^3} \frac {d^3\vec k}{(2\pi)^3}  \frac {b_3(\vec k)}{2\bar\omega v^0} \left( \omega_> \tens v-\tens k_> \right) e^{-i\tens k_> \cdot \tens x}\cr
& +ig \chi \omega_0^2 \sum_{a=1}^2 \sum_{i=1}^2 \int_{\mathbb R^3} \frac {d^3\vec k}{(2\pi)^3} \frac {a^{(a)}_i(\vec k)}{DR'_{(a)}} \frac {\omega_{(a)}}{\omega_{(a)}^2-\omega_0^2} \tens e_i^{(a)} e^{-i\tens k_{(a)} \cdot \tens x}\cr
& -ig \chi \omega_0^2 \int_{\mathbb R^3} \frac {d^3\vec k}{(2\pi)^3} \frac {a^\dagger_0(\vec k)}{2|\vec k|} (\omega_+ \tens v - \tens k_+)e^{i\tens k_+ \cdot \tens x}
-i \int_{\mathbb R^3} \frac {d^3\vec k}{(2\pi)^3}  \frac {b^\dagger_3(\vec k)}{2\bar\omega v^0} \left( \omega_> \tens v- \tens k_> \right) e^{i\tens k_> \cdot \tens x}\cr
& -ig \chi \omega_0^2 \sum_{a=1}^2 \sum_{i=1}^2 \int_{\mathbb R^3} \frac {d^3\vec k}{(2\pi)^3} \frac { a^{(a)\dagger}_i(\vec k)}{DR'_{(a)}} \frac {\omega_{(a)}}{\omega_{(a)}^2-\omega_0^2} \tens e_i^{(a)} e^{i\tens k_{(a)} \cdot \tens x},\\
B(\tens x) =& i \int_{\mathbb R^3} \frac {d^3\vec k}{(2\pi)^3} \frac {a_0(\vec k)}{2|\vec k|} \frac {\omega_+}{4\pi} (\omega_+^2-\bar\omega^2) e^{-i\tens k_+ \cdot \tens x}
 -i \int_{\mathbb R^3} \frac {d^3\vec k}{(2\pi)^3} \frac {a^\dagger_0(\vec k)}{2|\vec k|} \frac {\omega_+}{4\pi} (\omega_+^2-\bar\omega^2) e^{i\tens k_+ \cdot \tens x}.
\end{align}
As we see, the expressions for the fields now contain a divergence in $\omega=\omega_0$ intersected with $k^2=0$. We shall postpone the choice of the test functions until section \ref{sec: propagator}. However, it is worth noting that the divergences
appear exclusively in unphysical components of the $\tens A$ field. 
Here the dagger means complex conjugation. In order to represent the algebra of the canonical commutation relations, it is convenient to compute also the conjugate momenta. From \cite{PhysicaScripta}, and the above
construction, we get for the conjugate momenta the following expressions:
\begin{align}
\Pi^0_{\tens A}(\tens x)=& i \int_{\mathbb R^3} \frac {d^3\vec k}{(2\pi)^3} \frac {a_0(\vec k)}{2|\vec k|} \frac {\omega_+}{4\pi} (\omega_+^2-\bar\omega^2) e^{-i\tens k_+ \cdot \tens x}
 -i \int_{\mathbb R^3} \frac {d^3\vec k}{(2\pi)^3} \frac {a^\dagger_0(\vec k)}{2|\vec k|} \frac {\omega_+}{4\pi} (\omega_+^2-\bar\omega^2) e^{i\tens k_+ \cdot \tens x},\\
\Pi^l_{\tens A}(\tens x)=& i \int_{\mathbb R^3} \frac {d^3\vec k}{(2\pi)^3} \frac {a_0(\vec k)}{2|\vec k|}(k_+^0 v^l-k_+^l v^0)  \frac 1{4\pi}  (\omega_+^2-\bar\omega^2) e^{-i\tens k_+ \cdot \tens x}\cr
&+2i\sum_{a=1}^2 \sum_{j=1}^2 \int_{\mathbb R^3} \frac {d^3\vec k}{(2\pi)^3} \frac { a_j^{(a)}(\vec k)}{DR'_{(a)}} \tens e_j^{(a)[l} \left[ \frac 1{4\pi} k^{0]}_{(a)} -g^2 \frac {\chi \omega_0^2 \omega_{(a)}}{\omega_{(a)}^2
-\omega_0^2}v^{0]} \right] e^{-i\tens k_{(a)} \cdot \tens x} \cr
& -i \int_{\mathbb R^3} \frac {d^3\vec k}{(2\pi)^3} \frac {a^\dagger_0(\vec k)}{2|\vec k|}(k_+^0 v^l-k_+^l v^0) \frac 1{4\pi} (\omega_+^2-\bar\omega^2) e^{i\tens k_+ \cdot \tens x}\cr
&-2i\sum_{a=1}^2 \sum_{j=1}^2 \int_{\mathbb R^3} \frac {d^3\vec k}{(2\pi)^3} \frac { a_j^{(a)\dagger}(\vec k)}{DR'_{(a)}} \tens e_j^{(a)[l} \left[ \frac 1{4\pi} k^{0]}_{(a)} -g^2 \frac {\chi \omega_0^2 \omega_{(a)}}{\omega_{(a)}^2
-\omega_0^2}v^{0]} \right] e^{i\tens k_{(a)} \cdot \tens x},\\
\tens \Pi_{\tens P}(\tens x) =& -g v^0 \int_{\mathbb R^3} \frac {d^3\vec k}{(2\pi)^3} \frac {a_0(\vec k)}{2|\vec k|} \omega_+ (\omega_+ \tens v- \tens k_+) e^{-i\tens k_+ \cdot \tens x}
- \int_{\mathbb R^3} \frac {d^3\vec k}{(2\pi)^3} \frac {b_3(\vec k)}{2\chi \omega_0^2} (\omega_> \tens v-\tens k_>) e^{-i\tens k_> \cdot \tens x}\cr
& -g v^0 \sum_{a=1}^2 \sum_{j=1}^2\int_{\mathbb R^3} \frac {d^3\vec k}{(2\pi)^3} \frac {a_j^{(a)}(\vec k)}{DR'_{(a)}} \frac {\omega_{(a)}^2}{\omega_{(a)}^2-\omega_0^2} \tens e_j^{(a)} e^{-i\tens k_{(a)} \cdot \tens x}\cr
&-g v^0 \int_{\mathbb R^3} \frac {d^3\vec k}{(2\pi)^3} \frac {a^\dagger_0(\vec k)}{2|\vec k|} \omega_+ (\omega_+ \tens v- \tens k_+) e^{i\tens k_+ \cdot \tens x}
- \int_{\mathbb R^3} \frac {d^3\vec k}{(2\pi)^3} \frac {b^\dagger_3(\vec k)}{2\chi \omega_0^2} (\omega_> \tens v- \tens k_>) e^{i\tens k_> \cdot \tens x}\cr
& -g v^0 \sum_{a=1}^2 \sum_{j=1}^2\int_{\mathbb R^3} \frac {d^3\vec k}{(2\pi)^3} \frac {a_j^{(a)\dagger}(\vec k)}{DR'_{(a)}} \frac {\omega_{(a)}^2}{\omega_{(a)}^2-\omega_0^2} \tens e_j^{(a)} e^{i\tens k_{(a)} \cdot \tens x},\\
\Pi_B(\tens x)=& 0,
\end{align}
where the antisymmetrization $A^{[a}B^{b]}$ means $(A^a B^b-A^b B^a)/2$.
From these expressions we can extrapolate a basis of {\it quasi} plane wave solutions, which we collect into nine dimensional vectors of the form
\begin{eqnarray}
\pmb \zeta (\tens x)=
\begin{pmatrix}
\tens A(\tens x) \\ \tens P(\tens x) \\ B(\tens x)
\end{pmatrix}.
\end{eqnarray}
In particular, we define
\begin{align}
\pmb \zeta_0(\tens x,\vec k)=&
\begin{pmatrix}
(\omega_+^2-\omega_0^2)\tens v +\frac 12 \left( \frac \xi{4\pi} (\omega_+^2-\bar\omega^2)-\omega_+^2+\omega_0^2\right) (\tens v-i(\tens v \cdot \tens x)\tens k_+) -
\frac {i}{4\omega_+} Z(\vec k)\tens k_+\\[0.3cm]
ig {\chi \omega_0^2} (\omega_+ \tens v - \tens k_+) \\[0.3cm]
i\frac {\omega_+}{4\pi} (\omega_+^2-\bar\omega^2)
\end{pmatrix} e^{-i\tens k_+\cdot \tens x},\cr
\pmb \zeta_3(\tens x,\vec k)=&
\begin{pmatrix}
i\tens k_+ \\ \tens 0 \\ 0 
\end{pmatrix} e^{-i\tens k_+\cdot \tens x}, \cr 
\pmb {\tilde\zeta}_3(\tens x,\vec k)=&
\begin{pmatrix}
4\pi g  \left( \tens v- \frac {\bar \omega}{k_>^2} \tens k_>\right) \\[0.3cm] 
i(\omega_> \tens v-\tens k_>) \\ 0 
\end{pmatrix} e^{-i\tens k_>\cdot \tens x}, \cr
\pmb \zeta_{a,i}(\tens x,\vec k)=&
\begin{pmatrix}
\tens e_i^{(a)} \\[0.3cm]
ig \frac {\chi \omega_0^2 \omega_{(a)}}{\omega_{(a)}^2-\omega_0^2} \tens e_i^{(a)} \\[0.3cm]
0
\end{pmatrix} e^{-i\tens k_{(a)}\cdot \tens x}, \quad\ a=1,2,\ \ i=1,2,
\end{align}
where
\begin{eqnarray}
Z(\vec k):=\frac \xi{4\pi} (\omega_+^2-\bar\omega^2)+\omega_+^2-\omega_0^2  -\frac {
8\pi {\omega_+^2 g^2 \chi \omega_0^2}}{\omega_+^2-\bar\omega^2}.
\end{eqnarray}
The corresponding momenta are
\begin{align}
\pmb \Pi_{\pmb \zeta_0}(\tens x,\vec k)=&
\begin{pmatrix}
\begin{bmatrix}
i\frac 1{4\pi} (\omega_+^2-\bar\omega^2)\omega_+\\[0.3cm]
i \frac 1{4\pi} (\omega_+^2-\bar\omega^2) (k^0_+ \vec v- v^0\vec k_+) 
\end{bmatrix}\\[0.5cm]
-g v^0 { \omega_+} (\omega_+ \tens v - \tens k_+) \\[0.3cm]
0
\end{pmatrix} e^{-i\tens k_+\cdot \tens x},\qquad\
\pmb \Pi_{\pmb \zeta_3}(\tens x,\vec k)=
\begin{pmatrix}
\begin{bmatrix}
0 \\ \vec 0 
\end{bmatrix}\\ \tens 0 \\ 0 
\end{pmatrix}, \cr 
\pmb \Pi_{\pmb {\tilde\zeta}_3} (\tens x,\vec k)=&
\begin{pmatrix}
\begin{bmatrix}
0\\ \vec 0 
\end{bmatrix}\\[0.3cm]
- \frac {v^0\bar \omega}{\chi \omega_0^2}(\omega_> \tens v-\tens k_>) \\[0.3cm] 0 
\end{pmatrix} e^{-i\tens k_>\cdot \tens x}, \cr
\pmb \Pi_{\pmb \zeta_{a,i}}(\tens x,\vec k)=&
\begin{pmatrix}
\begin{bmatrix}
0\\
i\vec e_i^{(a)} \left[ \frac {k^0_{(a)}}{4\pi}-v^0 g^2 \frac {\chi \omega_0^2 \omega_{(a)}}{\omega_{(a)}^2-\omega_0^2} \right]- 
i e_i^{(a)0} \left[ \frac {\vec k_{(a)}}{4\pi}-\vec v g^2 \frac {\chi \omega_0^2 \omega_{(a)}}{\omega_{(a)}^2-\omega_0^2} \right] 
\end{bmatrix}\\
-g v^0 \frac { \omega_{(a)}^2}{\omega_{(a)}^2-\omega_0^2} \tens e_i^{(a)}\\[0.3cm]
0
\end{pmatrix} e^{-i\tens k_{(a)}\cdot \tens x}, \quad\ a,1=1,2.
\end{align}
In particular, let us compute the scalar products among the plane waves and the general solution. The conserved scalar product is \cite{PhysicaScripta}
\begin{eqnarray}
(\pmb {\tilde\zeta}| \pmb \zeta)=i\int_{\Sigma_t} d^3x \left(
\tilde A^{*\mu}(t;\vec x) \Pi_{\tens A \mu}(t;\vec x)+\tilde P^{*\mu}(t;\vec x) \Pi_{\tens P \mu}(t;\vec x)-\tilde \Pi_{\tens A}^{*\mu}(t;\vec x) A_\mu (t;\vec x)-\tilde \Pi_{\tens P}^{*\mu}(t;\vec x) P_\mu (t;\vec x)
\right),
\end{eqnarray}
where $\Sigma_t$ is any spacelike hypersurface. After some algebra we get 
\begin{align}
(\pmb \zeta_0(\vec k) | \pmb \zeta)=&-i\frac {\omega_+}{4\pi} (\omega^2-\bar\omega^2) a_3(\vec k),\\
(\pmb \zeta_3(\vec k) | \pmb \zeta)=& i\frac {\omega_+}{4\pi} (\omega^2-\bar\omega^2) a_0(\vec k), \\
(\pmb {\tilde\zeta}_3(\vec k) | \pmb \zeta)=& \frac {1}{\chi \omega_0^2} (\bar \omega^2- k_>^2) b_3(\vec k),\\
(\pmb \zeta_{a,i}(\vec k) | \pmb \zeta)=& a_i^{(a)} (\vec k).
\end{align}
Then, by using the inversions
\begin{align}
& a_0(\vec k)= -\frac {4\pi i}{\omega_+(\omega^2-\bar\omega^2)} (\pmb \zeta_3(\vec k) | \pmb \zeta),\\
& b_3(\vec k)= \frac {\chi \omega_0^2}{\bar \omega^2 - k_>^2} (\pmb {\tilde\zeta}_3(\vec k) | \pmb \zeta),\\
& a_i^{(a)}(\vec k)=(\pmb \zeta_{a,i}(\vec k) | \pmb \zeta),\\
& a_3(\vec k)=\frac {4\pi i}{\omega_+(\omega^2-\bar\omega^2)} (\pmb \zeta_0(\vec k) | \pmb \zeta),
\end{align}
we can compute the commutators among the Fourier modes.

\subsection{The quantum algebra of Fourier modes}
We take into consideration  the quantum algebra of fields, by promoting them to distributions taking value in a quantum algebra which realises the equal time canonical commutation relations among the fields $\tens A(\tens x)$, $\tens P(\tens x)$,
$B(\tens x)$, $\tens \Pi_{\tens A}(\tens x)$, $\tens \Pi_{\tens P}(\tens x)$, and $\Pi_B(\tens x)$, obtained by applying the correspondence principle to the respective Dirac brackets, see \cite{PhysicaScripta}. Since we need to rephrase the algebra
in terms of the particle representation (Fock representation), 
by means of the above formulas we are led to consider the commutators
\begin{eqnarray}
[(\pmb \zeta_A(\vec k)|\pmb \zeta), (\pmb \zeta_B(\vec q)|\pmb \zeta)], 
\end{eqnarray}
and
\begin{eqnarray}
[(\pmb \zeta_A(\vec k)|\pmb \zeta), (\pmb \zeta_B(\vec q)|\pmb \zeta)^*], 
\end{eqnarray}
where $A,B$ take the values $0,3,\tilde 3, (a,i)$, and $\pmb\zeta_{\tilde 3}=\pmb {\tilde\zeta}_3$. By using the definition of the scalar product in terms of the Poissonian structure, and the canonical commutation relations among the fields,
we get
\begin{eqnarray}
&& [(\pmb \zeta_A(\vec k)|\pmb \zeta), (\pmb \zeta_B(\vec q)|\pmb \zeta)]=(\pmb \zeta^*_B(\vec q)|\pmb \zeta_A(\vec k)), \label{c-zero}\\
&& [(\pmb \zeta_A(\vec k)|\pmb \zeta), (\pmb \zeta_B(\vec q)|\pmb \zeta)^*]=(\pmb \zeta_A(\vec k)|\pmb \zeta_B(\vec q)). \label{c-nonzero}
\end{eqnarray}
The scalar products (\ref{c-zero}) are obviously zero. This means that, if we set $\alpha_A$ so that $\alpha_0=a_0, \alpha_3=a_3, \alpha_{\tilde3}=b_3, \alpha_{a,i}=a_{a,i}$, then 
\begin{eqnarray}
[\alpha_A(\vec k), \alpha_B(\vec q)]=[\alpha_A^\dagger(\vec k), \alpha_B^\dagger(\vec q)]=0
\end{eqnarray}
as expected. On the other hand, we get 
\begin{align}
(\pmb\zeta_0(\vec k)|\pmb\zeta_0(\vec q))=&0,\\
(\pmb\zeta_0(\vec k)|\pmb\zeta_3(\vec q))=&-2i(2\pi)^3\frac {\omega_+}{4\pi} |\vec k| (\omega^2-\bar\omega^2) \delta^3(\vec k-\vec q),\\
(\pmb\zeta_3(\vec k)|\pmb\zeta_3(\vec q))=&0\\
(\pmb{\tilde\zeta}_3(\vec k)|\pmb{\tilde\zeta}_3(\vec q))=&2(2\pi)^3\frac {v^0\bar\omega}{\chi\omega_0^2} (\bar\omega^2-k_>^2)\delta^3(\vec k-\vec q),\\
(\pmb\zeta_{a,i}(\vec k)|\pmb\zeta_{b,j}(\vec q))=&(2\pi)^3 \left(\frac {k^0}{2\pi} +2\omega v^0 \frac {g^2\chi \omega_0^4}{(\omega^2-\omega_0^2)^2}\right) \delta_{ab}\delta_{ij} \delta^3(\vec k-\vec q),
\end{align}
whereas all the remaining terms trivially vanish. Then, the nontrivial commutators are
\begin{align}
[a_0(\vec k), a^\dagger_0(\vec q)]=&[a_3(\vec k),  a^\dagger_3(\vec q)]=0,\\
[a_0(\vec k), a^\dagger_3(\vec q)]=&(2\pi)^3\frac {|\vec k|}{\omega_+} \frac {8\pi i}{\omega^2-\bar\omega^2}\delta^3(\vec k-\vec q),\label{commutatore singolare}\\
[b_3(\vec k), b^\dagger_3(\vec q)]=& 2(2\pi)^3\frac {\chi \omega_0^2 v^0\bar \omega}{\bar \omega^2-k_>^2} \delta^3(\vec k-\vec q),\\
[ a^{(a)}_i(\vec k), a^{(b)\dagger}_j(\vec q)]=& (2\pi)^3 \left(\frac {k^0}{2\pi} +2\omega_{(a)} v^0 \frac {g^2\chi \omega_0^4}{(\omega_{(a)}^2-\omega_0^2)^2}\right) \delta_{ij} \delta_{ab} \delta^3(\vec k-\vec q).
\end{align}
Notice that the commutator (\ref{commutatore singolare}) is singular in $\omega=\bar\omega$. This is not merely a consequence of the singular redefinition of the fields introduced above. It is easy to see that the singularity would be appeared
even with the original definitions.

\subsection{The Hamiltonian}
We shall discuss later about the generators of the Noether symmetries related to the Poincar\'e group. However, a privileged role is played by the Hamiltonian, since, more than a symmetry, together with the canonical commutation relations
it defines the dynamics of the system. For this reason we anticipate here the determination of the Hamiltonian. Even though we have not yet introduced the construction of the Fock representation, being the latter one quite standard (apart from some
technical points) we think there is not any ambiguity in speaking about the normal ordering of the operators, with respect to what will be the vacuum state of the representation. Then, we shall turn on the discussion of the Hamiltonian after the explicit 
construction of the representation. Now, we assume the normal ordering to be understood. 
The computation of the Hamiltonian in terms of the particle modes is quite involved by the presence of the $\delta^{(1)}$ contributions. However, its determination can be notably simplified by means of the following
trick. 
The Hamiltonian can be computed directly, with the warning that the only terms which are problematic are the ones containing $a_0^\dagger a_0$. The direct computation gives
\begin{align}
H=&\sum_{j=1}^2 \sum_{a=1}^2 \int_{\mathbb R^3} \frac {d^3\vec k}{(2\pi)^3} \frac 1{DR'_{(a)}} a_j^{(a)\dagger} (\vec k) a_j^{(a)} (\vec k) {k^0_{(a)}}  +
 \int_{\mathbb R^3} \frac {d^3\vec k}{(2\pi)^3} \frac {\omega_>^2-k_>^2}{2v^0 \chi \omega_0^2 \bar \omega} b_3^\dagger(\vec k) b_3(\vec k) {k^0_>}\cr
 & + i\int_{\mathbb R^3} \frac {d^3\vec k}{(2\pi)^3} \frac {\omega_+}{8\pi} (\omega^2-\bar\omega^2) (a_3^\dagger(\vec k)a_0(\vec k)-a_0^\dagger(\vec k)a_3(\vec k))+\ldots
\end{align}
where the dots stay for the $a^\dagger_0 a_0$ terms, whose computation requires to handle the derivatives of the delta function and it is quite tricky. In order to determine them, let us consider first the mixed term
\begin{align}
H_{03}:=i\int_{\mathbb R^3} \frac {d^3\vec k}{(2\pi)^3} \frac {\omega_+}{8\pi} (\omega^2-\bar\omega^2)(a_3^\dagger(\vec k)a_0(\vec k)-a_0^\dagger(\vec k)a_3(\vec k)).
\end{align}
By using the commutation rules we first notice that (since, in particular, $[a_0(\vec k),a_0^\dagger(\vec k)]=0$) 
\begin{eqnarray}
[H,B(t,\vec x)]=[H_{03},B(t,\vec x)]=i\frac {\partial B(t,\vec x)}{\partial t},
\end{eqnarray}
as expected. Now, let us consider the gauge field
\begin{eqnarray}
\tens A_3(t,\vec x):=\int_{\mathbb R^3} \frac {d^3\vec k}{(2\pi)^3} \frac i{2|\vec k|} \tens k_+ \left[ a_3(\vec k) e^{-i \pmb k_+ \cdot \pmb x} -a^\dagger_3(\vec k) e^{i \pmb k_+ \cdot \pmb x} \right],
\end{eqnarray}
and set
\begin{eqnarray}
H_{00}:=\int_{\mathbb R^3} \frac {d^3\vec k}{(2\pi)^3} f(\vec k) a_0^\dagger (\vec k) a_0(\vec k)
\end{eqnarray}
for the $a_0^\dagger a_0$ terms in the Hamiltonian. Thus, $f(\vec k)$ is completely fixed by requiring that
\begin{eqnarray}
[H,\tens A_3(t,\vec x)]=[H_{03}+H_{00},\tens A_3(t,\vec x)]\equiv i\frac {\partial \tens A_3(t,\vec x)}{\partial t},\label{commutatore-H03-A3}
\end{eqnarray}
when $\tens A_3$ is a solution, which means when all the particle modes, but $a_3$, are set to zero. One could be tempted to write $H_{00}=0$, but it would be wrong. 
If we look at the first line of (\ref{campo A}), we see that $H_{03}$ alone does not provide the right temporal evolution of the $a_0$ components and we must
take $f\neq 0$. Indeed, the $H_{00}$ term does not commute with $\tens A_3$, but, being the commutator proportional to $a_0$, this does not affects the relation (\ref{commutatore-H03-A3}), which must be true on shell 
(i.e. when $\tens A=\tens A_3$ is a solution, thus corresponding, in particular, to $a_0=0$). In order to determine $f$
we must impose the right behaviour for the commutator between $H$ and $A$. Imposing it to the solutions corresponding to have only $a_0$ non zero, this provides
\begin{eqnarray}
f(\vec k)=-\frac {\omega_+ v^0}{16\pi|\vec k|}\left(\frac \xi{4\pi} (\omega^2-\bar\omega^2)-\omega^2+\omega_0^2\right) (\omega^2-\bar\omega^2).
\end{eqnarray}
Thus, the Hamiltonian takes the form
\begin{align}
H=&\sum_{j=1}^2 \sum_{a=1}^2 \int_{\mathbb R^3} \frac {d^3\vec k}{(2\pi)^3} \frac 1{DR'_{(a)}} a_j^{(a)\dagger} (\vec k) a_j^{(a)} (\vec k) {k^0_{(a)}}  +
 \int_{\mathbb R^3} \frac {d^3\vec k}{(2\pi)^3} \frac {\bar\omega^2-k_>^2}{2v^0 \chi \omega_0^2 \bar \omega} b_3^\dagger(\vec k) b_3(\vec k) {k^0_>}\cr
 & + i\int_{\mathbb R^3} \frac {d^3\vec k}{(2\pi)^3} \frac {\omega_+}{8\pi} (\omega^2-\bar\omega^2)(a_3^\dagger(\vec k)a_0(\vec k)-a_0^\dagger(\vec k) a_3(\vec k))\cr
 &-\int_{\mathbb R^3}  \frac {d^3\vec k}{(2\pi)^3}  \frac {\omega_+ v^0}{16\pi |\vec k|}\left(\frac \xi{4\pi} (\omega^2-\bar\omega^2)-\omega^2+\omega_0^2\right) (\omega^2-\bar\omega^2)
 a_0^\dagger(\vec k)a_0(\vec k),
\end{align}
which, indeed, provides the right time evolution of all the fields.
Notice the compatibility with the commutation rules. A direct comparison shows that we can interpret 
\begin{eqnarray}
dn^{(a)}_j (\vec k)=\frac 1{DR'_{(a)}}  a_j^{(a)\dagger} (\vec k)  a_j^{(a)} (\vec k) \frac {d^3\vec k}{(2\pi)^3}
\end{eqnarray}
as the number of excitations of the field, in the mode $\zeta^{(a)}_i$ with linear momentum $\vec k$ in a volume $d^3\vec k$. This should be corroborated by the explicit form of the linear momentum operator.
Analogously 
\begin{eqnarray}
dn_{b_3}(\vec k)=\frac {\bar\omega^2-k_>^2}{2v^0 \chi \omega_0^2 \bar \omega} b_3^\dagger(\vec k) b_3(\vec k)  \frac {d^3\vec k}{(2\pi)^3}
\end{eqnarray}
is the number of states for the excitation of the $\omega_> \tens v- \tens k_>$ component of the field, with linear momentum $\vec k$ in a volume $d^3\vec k$. We have intentionally avoided to speak about excitation of the electromagnetic
field rather than the polarisation field, in the light of the fact that the excited modes indeed correspond to 
an excitation of both $\tens A$ and $\tens P$. This is the effect of the ``interaction'' in the Heisenberg representation: fields 
$\tens A$ and $\tens P$ are not associated with independent particle modes (of course, this would happen in the interaction representation).  

\section{Covariance of the theory under the Poincar\'e group and causality}\label{sec:covariance and causality}
Let us now discuss the covariance properties of the theory constructed up to now. The classical theory is covariant since it provides a working model in any specified inertial frame, selected by a specification of the velocity vector
$\tens v$. Indeed, it is immediate to verify that, consistently, given a Lorentz transformation $\tens \Lambda \in SO(1,3)$ such that $\tens v'=\tens \Lambda \tens v$, then the theory specified by $\tens v'$ is obtained by transforming 
the system specified by $\tens v$ under $\tens \Lambda$. Nevertheless, covariance is not the same as invariance: the presence of the dielectric medium, and then of the vector $\tens v$, explicitely breaks the Poincar\'e symmetry so that, in general, 
boosts are no more expected to be symmetries, and, indeed, they are not. For example, if we consider the free oscillators describing the dielectric medium, thus with coupling $g=0$, its equations of motion in the frame at rest w.r.t. the medium are
\begin{align}
\partial_t^2 P^i+\omega_0^2 P^i=&0,\\
P^0=&0.
\end{align}
Their covariant form is 
\begin{align}
(v^\nu \partial_\nu)^2 P^\mu+\omega_0^2 P^\mu=&0, \\
v^\mu P_\mu=&0.
\end{align}
The Lorentz symmetry is broken since one can individuate the inertial frame just by looking at the form of the equations of motion. Nevertheless, the equations remain covariant in the sense that the first set of equations
can be recovered from the second one simply by setting $\tens v \equiv (1;\vec 0)$, and in any other frame by specifying the velocity $\tens v$.\\
In order to discuss this point appropriately, as a general fact, let us consider a theory for a field $\psi$ in which a velocity $\tens v$, specifying a privileged inertial frame, 
appears explicitly in the Lagrangian in such a way that the theory in a given frame, boosted by $\vec v/v^0$ with respect to the privileged frame, is described simply by fixing the spacetime vector $\tens  v\equiv (v^0,\vec v)$. 
Obviously, our Lagrangian belongs into this class. The point is that such a Lagrangian \underline{is not} a scalar under Lorentz transformations unless we transform $\tens v$ also. In other words we have
\begin{eqnarray}
L(\psi(\tens\Lambda \tens x), \partial_\mu \psi(\tens\Lambda \tens x); \tens \Lambda \tens v)=L(\psi( \tens x), \partial_\mu \psi( \tens x); \tens v),
\end{eqnarray}
whereas
\begin{eqnarray}
L(\psi(\tens \Lambda \tens x), \partial_\mu \psi(\tens\Lambda \tens x); \tens v)\neq L(\psi(\tens x), \partial_\mu \psi(\tens x); \tens v).
\end{eqnarray}
Thus, such a transformation does not represents a symmetry of a specified model, but it is indeed a covariance transformation and, technically, it looks more as a duality transformation relating two different set of parameters ($\tens v$ and $\tens v'$) 
in a family of models. In order to further clarify this point, let us investigate quickly its role in the Noether theorem.
If we consider a Lorentz transformation $\tens \Lambda$ specified by an infinitesimal transform $\epsilon$ such that $\eta\epsilon=-\epsilon^T  \eta$, $\eta$ being the usual Minkowski tensor, then the field will transform as
\begin{eqnarray}
\delta_\epsilon \psi(\tens x)= \frac 12 \epsilon_{\mu\nu}\sigma^{\mu\nu} \psi(\tens x)+\epsilon^{\mu\nu}x_\mu \partial_\nu \psi(\tens x),
\end{eqnarray}
where $\sigma^{\mu\nu}$ are the representation matrices of the Lorentz group determined by the spin of the field. Such a transformation, as discussed above, must be also associated with a Lorentz transformation of the velocity
\begin{eqnarray}
v^\mu \longmapsto v^\mu+\epsilon^\mu_{\ \nu} v^\nu
\end{eqnarray}
for consistency. Only after including this transformation, the Lagrangian will transform as a scalar, which means
\begin{eqnarray}
\frac {\partial L}{\partial \psi} \delta_\epsilon \psi +\frac {\partial L}{\partial \partial_\mu\psi} \partial_\mu\delta_\epsilon \psi +\frac {\partial L}{\partial v^\mu} \delta_\epsilon v^\mu= \epsilon^{\mu\nu}x_\mu \partial_\nu L(\tens x),
\end{eqnarray}
where in the r.h.s. we emphasised that the Lagrangian has been seen as a scalar function of $\tens x$. Equivalently, this can be recast in the form
\begin{eqnarray}
\left( \frac {\partial L}{\partial \psi}- \partial_\mu \frac {\partial L}{\partial \partial_\mu\psi} \right) \delta_\epsilon \psi +\partial_\mu \left(\frac {\partial L}{\partial \partial_\mu\psi} \delta_\epsilon \psi +\epsilon^{\mu\nu}x_\nu L\right)
=-\frac {\partial L}{\partial v^\mu} \delta_\epsilon v^\mu. \label{Noether}
\end{eqnarray}
Usually, for theories with scalar couplings, the right hand side vanishes so that on any solution of the equations of motion the quantity
\begin{eqnarray}
j_\epsilon^\mu(\tens x)=\frac {\partial L}{\partial \partial_\mu\psi} \delta_\epsilon \psi +\epsilon^{\mu\nu}x_\nu L
\end{eqnarray}
defines a conserved current. But in our general case the story is quite different since the r.h.s. of (\ref{Noether}) is different from zero. One may wonder if $\frac {\partial L}{\partial v^\mu} \delta_\epsilon v^\mu$ is a divergence, but also this
cannot be true in general. For example, if $L$ depends polynomially on $\tens v$, like in our specific case, then this would imply that also each term in $L$ explicitly containing $\tens v$ would be a divergence. But in this case such terms would 
not influence the dynamics and could be completely eliminated from the action from the beginning.\\
Thus, the Lorentz symmetry is broken down to one of its maximal compact subgroups: the little group associated to $\tens v$, which, $\tens v$ being a timelike vector, is a three dimensional rotations group. Since it leaves $\tens v$
invariant, we have that, for $\epsilon_{\tens v}$ a generator of the little group, $\delta_{\epsilon_{\tens v}} v^\mu=0$ and the corresponding current $\tens j_{\epsilon_{\tens v}}$ is conserved.
In other words, the presence of the dielectric medium defines a sort of external source which does not allow for the conservation of the ergocenter velocity (and maybe the ergocenter is not even well defined since the dielectric medium essentially looks like 
a system with infinite inertia).\\
We can rephrase the problem we discussed above also by referring to the ideas developed in \cite{anderson}. See also \cite{sundermeyer}, pp. 356-360. 
Our model Lagrangian is associated with equations of motion which are covariant with respect to the Lorentz group. This fact does not correspond to a full Lorentz invariance, due to the fact that the Lorentzian metric is not the only absolute object of the theory, 
but the velocity $\tens v$ of the medium is an absolute object too. This implies that our theory, and any covariant theory of a dielectric medium (cf. e.g. \cite{jauch1}), is involved with a preferred frame: the rest frame of the medium. This kind of behaviour is not 
exceptional, as it occurs e.g. in all the cases where Klein-Gordon equation is studied in presence of an external potential. Loss of Lorentz invariance is evident, but, at the same time, the field equations are covariant, and solutions are transformed into solutions of 
the Klein-Gordon equation by the Lorentz group, provided that external potential is transformed 
too \cite{gerard}.\\

We shall see soon how to compute the conserved charges. 
Before doing this, we want to discuss the covariance of the quantum theory on the light of the stated observations. 
\subsection{Covariance and the quantum theory} How do the above considerations reflect in the quantum theory construction? Since the Lorentz group is no more a symmetry group, but only the little group of $\tens v$ for any given $\tens v$,
beyond spacetime translations, we cannot expect to represent the whole Poincar\'e group on a single Fock space, but over a family $\mathcal F_{\tens v}$ of Fock spaces, one for each $\tens v$.  Each $\mathcal F_{\tens v}$ will be
cyclically generated by the polynomial action of the fields on a vacuum state $|0\rangle_{\pmb v}$, annihilated by the set of operators 
\begin{eqnarray}
a_{0}(\vec k;\tens v),\  a_{3}(\vec k;\tens v),\ b_{3}(\vec k;\tens v),\ a^{(a)}_{i}(\vec k;\tens v),\ i=1,2,\ a=1,2,\ \vec k\in \mathbb R^3 \label{pmb v operators}.
\end{eqnarray}
We shall refer to the Fock space $\mathcal F_{\tens v}$ as the $\tens v$ sector. Since they correspond to different inertial frames, any two sectors are expected to be equivalent and, then, related by a boost in some way. 
More precisely, we infer the existence of a map
\begin{align}
U: SO(1,3)^{\uparrow}\times V^+ &\longrightarrow  \mathcal {U}, \\
 (\tens\Lambda,\tens v)  &\longmapsto  U(\tens \Lambda,\tens v) : \mathcal {F}_{\tens v} \longrightarrow \mathcal {F}_{\tens \Lambda \tens v}, \ \forall \tens v \text{ timelike},
\end{align}
from the orthocronus Lorentz group times the set of all future directed timelike vectors to the set $\mathcal {U}$ of all unitary maps among Fock spaces $\mathcal {F}_{\tens v}$ and $\mathcal {F}_{\tens w}$, in such a way that $U(\tens \Lambda,\tens v)$ 
must satisfy
\begin{eqnarray}
U(\tens \Lambda,\tens v)|0\rangle_{\tens v}=|0\rangle_{\tens \Lambda \tens v}, \label{U su vuoto}
\end{eqnarray}
and
\begin{eqnarray}
U(\tens\Lambda, \tens v) o(\vec k; \tens v) U(\tens\Lambda, \tens v)^{-1}=o(\tens\Lambda\vec k; \tens \Lambda\tens v), \label{U su operatori}
\end{eqnarray}
for $o$ any of the operators in (\ref{pmb v operators}). Here, with $\tens\Lambda\vec k$ we mean $\tens \Lambda(k^0(\vec k), \vec k)$, which makes perfectly sense since all dispersion relations are Lorentz covariant.
The maps $U(\tens \Lambda,\tens v)$ are indeed well defined and unitary. There is no particular difficulty in proving that 
and we shall sketch the proof.\\
Given $\tens \Lambda \in SO(1,3)^{\uparrow}$, $\tens v\in V^+$, and setting $\tens w=\tens \Lambda \tens v$, we can construct the Fock spaces $\mathcal F_{\tens v}$ and $\mathcal F_{\tens w}$, in the usual way, as the Cauchy completion of the
spaces generated by the polynomial actions of the operators
\begin{eqnarray}
\int_{\mathbb R^3} d^3\nu_o(\vec k) o^\dagger(\vec k, \tens z) f(\vec k),
\end{eqnarray}
on $|0\rangle_{\tens z}$, where $\tens z=\tens v, \tens w$, $o$ is any of the operators (\ref{pmb v operators}), $d^3\nu_a(\vec k)$ the corresponding covariant measure,\footnote{so
\begin{eqnarray}
d^3\nu_o(\vec k)=\frac {d^3\vec k}{(2\pi)^3 DR_o'(\vec k)},\label{measure}
\end{eqnarray}
where $DR_o=0$ is the dispersion relation describing the $\tens k$ spectrum relative to the operator $o$, and the prime means derivative w.r.t. $k^0$.
} and $f$ is any test function (wave packet). This way, the two Fock spaces are well defined, and the maps defined by (\ref{U su vuoto}) and (\ref{U su operatori}) are also well defined and isometric on all polynomially generated states (obviously
one has to take into account all subtleties relative to the restriction to the set of physical states, projecting out the  null norm states, but it is standard so we do not enter into details). Since the maps just defined are bounded operators over dense sets,
they remain well defined by Cauchy completion, and provide indeed unitary equivalences among the different representations.

\

In other words, the representation of the quantum algebra is finally realised on a bundle of Fock spaces over the quotient space $\mathcal B=SO(1,3)^{\uparrow}/SO(3)\simeq V^+$. The detailed construction of the Fock spaces (and, then, the details of the proof sketched above) is omitted as it is standard.

\subsection{On the causality properties}
In a quantum field theory, causality imposes that the commutators among fields vanish when the supports of the fields have a spacelike separation, or, in the point-like formulation, if the fields are evaluated at spatially separated points.
Having ensured covariance, in our construction such causality relations are simply a consequence of the equal time canonical commutation relations. Notice that in the presence of gauge fields this cannot be realised when also charged fields are involved.
However, in our case no charged fields are present.

\subsection{Conserved quantities and the little group}
For completeness let us shortly discuss the conserved quantities associated to the continuous symmetries of the relativistic Hopfield model. The symmetry group is $G_{\tens v} \ltimes \mathbb R^4$, the semidirect product of the little group $G_{\tens v}$
of $\tens v$ and the spacetime translations group $\mathbb R^4$. The conserved currents associated with translations do not require any particular care, and are computed exactly in the same way as for any relativistic invariant theory. The same happens
for the little group, with the only care to appropriately select the correct infinitesimal transformations. For a Lorentz invariant theory, the infinitesimal generators are parameterised by any antisymmetric matrix $\epsilon_{\mu\nu}$, and give rise to six
conserved quantities $M^{\mu\nu}=-M^{\nu\mu}$, which, for a set of fields $\tens \Psi_a$, $a=1,\ldots,n$, having spin $S_a$ (with representation matrices $\pmb\sigma^{\mu\nu}=-\pmb\sigma^{\nu\mu}$) and Lagrangian $\mathcal L$, can be expressed 
in the form
\begin{eqnarray}
K^j=M^{0j}=\sum_{a=1}^n\int_\Sigma [i\pmb\sigma^{0j} \tens \Psi_a(\tens x) \pi_\alpha(\tens x)+(x^0\partial^j-x^j\partial^0)\tens\Psi_a(\tens x) \pi_a(\tens x)+\frac 1n x^j \mathcal L]d^3\vec x
\end{eqnarray}
for the boosts, where $\pi_a$ is the momentum field canonically conjugate to $\tens \Psi_a$, whereas for rotations we have 
\begin{eqnarray}
J_i=\frac 12 \epsilon_{ijk}M^{jk}=\sum_{a=1}^n\int_\Sigma [i{\tens S_i}\tens \Psi_a(\tens x) \pi_a(\tens x)+\frac 12 \epsilon_{ijk} (x^j\partial^k-x^k\partial^j)\tens \Psi_a(\tens x) \pi_a(\tens x)]d^3\vec x,
\end{eqnarray}   
with $2\tens S_i=\epsilon_{ijk}\pmb \sigma^{jk}$. However, we cannot apply directly this formula to our case since we have to restrict the parameters generating symmetries to the set of antisymmetric matrices $\epsilon_{\mu\nu}$ that satisfy
\begin{eqnarray}
\epsilon_{\mu\nu} v^\mu=0.
\end{eqnarray}
The solutions are given by 
\begin{eqnarray}
\epsilon_{\mu\nu}=\varepsilon_{ij} \tens \Lambda(\tens v)^i_{\ \mu}\tens \Lambda(\tens v)^j_{\ \nu},
\end{eqnarray}
where $\varepsilon_{\mu\nu}$ is an arbitrary antisymmetric matrix which vanishes if one of the indices is zero, and $\tens\Lambda(\tens v)$ is any boost transforming $\tens v$ into $\tens c\equiv (1,\vec 0)$.
We can choose
\begin{align}
\tens \Lambda(\tens v)^0_{\ 0}&= {v^0},\\
\tens\Lambda(\tens v)^i_{\ 0}&=\tens\Lambda(\tens v)^0_{\ i}=-v^i,\\
\tens\Lambda(\tens v)^i_{\ j}&=(\gamma(\tens v)-1) \frac {v^iv^j}{\vec v^2}+\delta^i_j.
\end{align}
In this way we see that the little group rotations are generated by the conserved quantities
\begin{eqnarray}
R^{ij}=\tens\Lambda(\tens v)^i_{\ \mu} \tens\Lambda(\tens v)^j_{\ \nu} M^{\mu\nu},
\end{eqnarray}
where $M^{\mu\nu}$ are the quantities defined above (and computed in our specific case). More precisely
\begin{eqnarray}
R^{ij}=M^{ij}-(\gamma(\tens v)-1) \frac {v_k}{\vec v^2} (v^i M^{kj}-v^j M^{ki})- v^i M^{0j}+v^j M^{0i},
\end{eqnarray}
where $v_k=-v^k$, or, equivalently, recalling that $\epsilon_{ijk}=-\epsilon^{ijk}$:
\begin{eqnarray}
J_{\tens v}^l:=\frac 12 \epsilon^{lij} R_{ij}=v^0 J^l-\epsilon^{lij} v_i K_j,
\end{eqnarray}
with $J^l$ and $K^l$ as given above. Obviously, the normal ordering w.r.t. the vacuum state $|0\rangle_v$ has to be understood in the quantum expressions. We don't display the explicit formulas giving the operators
$J^l$ in terms of the annihilation and creation operators, as they are cumbersome and of little interest. \\

\section{The two points function}\label{sec: propagator}
Our theory, being quadratic in the fields, is essentially characterised by the propagator, which can be computed directly as the vacuum expectation value of the time ordered product of the fields
\begin{eqnarray}
iG_{\tens v}^{IJ}(\tens x,\tens y)=   {}_{\tens v}\langle 0| T(\Phi^I(\tens x)\Phi^J(\tens y))|0\rangle_{\tens v}, \qquad I,J=1,\ldots,9,
\end{eqnarray}
where 
\begin{eqnarray}
\Phi^I=
\begin{cases}
A^{I-1} & \mbox{if $I=1,2,3,4$}, \\
P^{I-5} & \mbox{if $I=5,6,7,8$}, \\
B & \mbox{if $I=9$}.
\end{cases}
\end{eqnarray}
We shall also write
\begin{eqnarray}
G_{\tens v}^{IJ}(\tens x,\tens y)=G_{\tens v}^{IJ}(\tens x,\tens y)_+\theta (x^0-y^0)+G_{\tens v}^{IJ}(\tens x,\tens y)_-\theta(y^0-x^0).
\end{eqnarray}
Since $G_{\tens v}^{IJ}(\tens x, \tens y)_-$ is easily obtained from $G_{\tens v}^{IJ}(\tens x,\tens y)_+$, we shall write down only the latter. This computation does not require any particular shrewdness and we can directly write down the relevant
components, which are
\begin{align}
iG_{\tens v}^{(\mu+1)(\nu+1)}(\tens x,\tens y)_+=&\ {}_{\tens v}\langle 0| A^\mu(\tens x)A^\nu(\tens y)|0\rangle_{\tens v}\cr
=&\int_{\mathbb R^3} \frac {d^3\vec k}{(2\pi)^3}e^{-i\tens k_+\cdot (\tens x-\tens y)}
\frac {v^\mu k_+^\nu+v^\nu k_+^\mu}{4|\vec k| \omega_+} \left(\xi+\frac  {4\pi (\omega^2-\omega_0^2)}{\omega^2-\bar\omega^2}  \right)\cr
&-i \int_{\mathbb R^3} \frac {d^3\vec k}{(2\pi)^3}e^{-i\tens k_+\cdot (\tens x-\tens y)}\frac {(\tens x-\tens y)\cdot \tens v}{4|\vec k|\omega_+}\left(\xi-\frac  {4\pi(\omega^2-\omega_0^2)}{(\omega^2-\bar\omega^2)}\right) k^\mu k^\nu\cr
&+ \int_{\mathbb R^3} \frac {d^3\vec k}{(2\pi)^3}e^{-i\tens k_+\cdot (\tens x-\tens y)} \pi  \frac {\frac \xi{4\pi}(\omega^2-\bar\omega^2)^2+\left(\omega^2-\omega_0^2\right)(\omega^2-\bar\omega^2)
-{8\pi\omega_+^2 g^2 \chi \omega_0^2}}{\omega_+^2 |\vec k| (\omega^2-\bar\omega^2)^2}
k^\mu k^\nu\cr
&+\frac {8\pi^2 g^2 \chi \omega_0^2}{\bar \omega v^0} \int_{\mathbb R^3} \frac {d^3\vec k}{(2\pi)^3}e^{-i\tens k_>\cdot (\tens x-\tens y)}\frac 1{\bar\omega^2- k^2_>} \left( v^\mu-\frac {\bar\omega}{k^2_>} k^\mu_>\right)
\left( v^\nu-\frac {\bar\omega}{k^2_>} k^\nu_>\right)\cr
&+\sum_{a=1}^2 \sum_{i=1}^2 \int_{\mathbb R^3} \frac {d^3\vec k}{(2\pi)^3}e^{-i\tens k_{(a)}\cdot (\tens x-\tens y)} \frac {e^{(a)\mu}_i(\vec k) e^{(a)\nu}_i (\vec k)}{DR'_{(a)}(\vec k)}; \label{AA}
\end{align}
\begin{align}
iG_{\tens v}^{(\mu+1)(\nu+5)}(\tens x,\tens y)_+=&\ {}_{\tens v}\langle 0| A^\mu(\tens x)P^\nu(\tens y)|0\rangle_{\tens v}\cr
=&-i \int_{\mathbb R^3} \frac {d^3\vec k}{(2\pi)^3}e^{-i\tens k_+\cdot (\tens x-\tens y)} \frac {2\pi g\chi \omega_0^2}{|\vec k| \omega_+} \frac {\omega_+ k_+^\mu v^\nu- k_+^\mu k_+^\nu}{\omega_+^2-\bar\omega^2} \cr
&-i\frac {2\pi g\omega_0^2 \chi}{\bar \omega v^0} \int_{\mathbb R^3} \frac {d^3\vec k}{(2\pi)^3}e^{-i\tens k_>\cdot (\tens x-\tens y)} \frac 1{\bar \omega^2-k_>^2} \left( v^\mu-\frac {\bar \omega}{k^2_>} k^\mu_> \right) (\bar \omega v^\nu- k_>^\nu)\cr
& -ig \chi \omega_0^2  \sum_{a=1}^2 \sum_{i=1}^2 \int_{\mathbb R^3} \frac {d^3\vec k}{(2\pi)^3}e^{-i\tens k_{(a)}\cdot (\tens x-\tens y)} \frac {\omega_{(a)}}{\omega_{(a)}^2-\omega_0^2} 
\frac {e^{(a)\mu}_i(\vec k) e^{(a)\nu}_i (\vec k)}{DR'_{(a)}(\vec k)}; \label{AP}
\end{align}
\begin{align}
iG_{\tens v}^{(\mu+1)9}(\tens x,\tens y)_+=&\ {}_{\tens v}\langle 0| A^\mu(\tens x)B(\tens y)|0\rangle_{\tens v}=-\frac i2 \int_{\mathbb R^3} \frac {d^3\vec k}{(2\pi)^3}e^{-i\tens  k_+\cdot (\tens x-\tens y)} \frac {k^\mu}{|\vec k|}; \label{AB}
\end{align}
\begin{align}
iG_{\tens v}^{(\mu+5)(\nu+5)}(\tens x,\tens y)_+=& {}_{\tens v}\langle 0| P^\mu(\tens x)P^\nu(\tens y)|0\rangle_{\tens v}\cr
=&\frac {\chi \omega_0^2}{2 \bar\omega v^0} \int_{\mathbb R^3} \frac {d^3\vec k}{(2\pi)^3}e^{-i\tens k_>\cdot (\tens x-\tens y)} \frac 1{\bar \omega^2-\omega_0^2} (\bar \omega v^\mu- k_>^\mu)(\bar \omega v^\nu- k_>^\nu)\cr
& +g^2 \chi^2 \omega_0^4  \sum_{a=1}^2 \sum_{i=1}^2 \int_{\mathbb R^3} \frac {d^3\vec k}{(2\pi)^3}e^{-i\tens k_{(a)}\cdot (\tens x-\tens y)} \frac {\omega_{(a)}}{\omega_{(a)}^2-\omega_0^2} 
\frac {e^{(a)\mu}_i(\vec k) e^{(a)\nu}_i (\vec k)}{DR'_{(a)}(\vec k)}; \label{PP}
\end{align}
\begin{align}
iG_{\tens v}^{(\mu+5)9}(\tens x,\tens y)_+=&\ {}_{\tens v}\langle 0| P^\mu(\tens x)B(\tens y)|0\rangle_{\tens v}=0; \label{PB}
\end{align}
\begin{align}
iG_{\tens v}^{99}(\tens x,\tens y)_+=&\ {}_{\tens v}\langle 0| B(\tens x)B(\tens y)|0\rangle_{\tens v}=0. \label{BB}
\end{align}
At a first sight, these quite involved expression do not immediately communicate any particular information. However, in the definition of the fields, we have seen that we need to restrict the space of test functions in order to avoid a singular behaviour around 
$\omega^2=\bar\omega^2$ intersected with the future directed light cone. By looking at the expressions for the propagator, we see precisely what is the degree at which the singularities appear. These new singularities have been originated by the 
commutator (\ref{commutatore singolare}) between the $a_0$ and the $a_3$ modes. In particular, it involves only gauge configuration and is thus expected
to be eliminable from any contribution to physical quantities. We can avoid them by a suitable choice of the distribution space, thus making all expressions well defined.

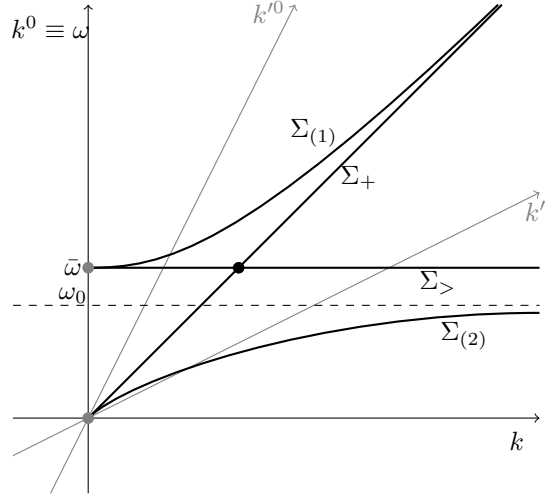
\begin{figure}[!thbp]
\begin{center}
\begin{tikzpicture} 
\draw [->](4,0) -- (11,0);
\draw [->](5,-1) -- (5,5.5);
\draw[gray] [->](4,-0.5) -- (11,3);
\draw[gray] [->](4.5,-1) -- (7.75,5.5);
\draw [-,dashed](4,1.5) -- (11,1.5);
\draw[-,thick](5,2)--(11,2);
\draw[thick] [-](5,0) -- (10.5,5.5);	
\draw[thick] (5,2) .. controls (6,2) and (7,2.2) .. (10.45,5.5);
\draw[thick] (5,0) .. controls (5.5,0.5) and (8,1.4) .. (11,1.4);
\node at (4.8,1.63) {$\omega_0$};
\node at (4.8,2) {$\bar\omega$};
\node at (4.5,5.2) {$k^0\equiv \omega$};
\node at (10.7,-0.3) {$k$};
\node[gray] at (7.4,5.4) {$k^{\prime0}$};
\node[gray] at (10.95,2.75) {$k'$};
\filldraw [gray]
(5,0) circle (2pt)
(5,2) circle (2pt);
\filldraw
(7,2) circle (2pt);
\node at (9.65,1.8) {$\Sigma_>$};
\node at (8.6,3.2) {$\Sigma_+$};
\node at (8,3.8) {$\Sigma_{(1)}$};
\node at (10,1.1) {$\Sigma_{(2)}$};
\end{tikzpicture}
\caption{The thick black lines represent the dispersion relations (as seen in the lab frame). The grey lines represent the axes of a boosted frame. The grey points are single point intersections. The black point is a two-dimensional
intersection surface.}\label{figura dispersion}
\end{center}
\end{figure}

The origin of the extra singularities is due to the fact that the different branches of the spectrum have nontrivial intersections. The three main components of the spectrum are defined respectively by the equations
\begin{align}
& k^2=0, \\
& \omega^2-\bar \omega^2=0, \\
& \frac {k^2}{4\pi} -\frac {g^2\chi_0 \omega_0^2 \omega^2}{\omega^2-\omega_0^2}=0,
\end{align}
together with positivity conditions, and which we shall indicate with $\Sigma_+$, $\Sigma_>$ and $\Sigma_{(a)}$ respectively. 
The denominators that generate divergences in the propagator have the form $(\omega_+^2-\bar \omega^2)^l$, $l=1,2$. The polar region is thus in the intersection $\Sigma_p=\Sigma_+ \cap \Sigma_>$. Since $k^0=|\vec k|$
this is equivalent to write
\begin{eqnarray}
|\vec k| v^0- \vec v\cdot \vec k=\pm \bar\omega.
\end{eqnarray}
Now $\pmb v^2=c^2$ implies $v^0>|\vec v|$ so that we must take the positive sign since $v^0$ is positive. The equation is thus equivalent to
\begin{eqnarray}
v_0^2 \vec k^2=(\bar \omega+\vec v \cdot \vec k)^2.\label{ellissoide}
\end{eqnarray}
If we now introduce the shift
\begin{eqnarray}
\vec k=\vec q+ \bar\omega \vec v,
\end{eqnarray}
the equation becomes
\begin{eqnarray}
\frac {\vec q^2}{\bar\omega^2} -  \frac {(\vec v \cdot \vec q)^2}{\bar\omega^2v_0^2}=1.
\end{eqnarray}
This shows that the singular locus is a prolate ellipsoid, symmetric under rotation around the direction of $\vec v$, centred in 
\begin{eqnarray}
\vec k=\bar\omega \gamma(\vec \nu) \vec \nu,
\end{eqnarray}
where
\begin{eqnarray}
\vec v=\gamma(\vec \nu) \vec \nu,\qquad\ \gamma(\vec \nu)=\frac 1{\sqrt {1-\vec \nu^2}},
\end{eqnarray}
with shorter axis of length 
\begin{eqnarray}
b=\bar\omega
\end{eqnarray}
whereas the longer one has length 
\begin{eqnarray}
b=\bar\omega \gamma(\vec \nu).
\end{eqnarray}
We can cure the singularities by defining the field operators as distributions acting on test rapidly decreasing smooth functions whose Fourier transforms vanish on
the singular ellipsoid at least with degree two. It is easily verified that this domain gives rise to well defined distributions. \\
It is interesting to note that the remaining intersections among spectra are just pointlike and give no new singularities. Indeed, we have that $\Sigma_+\cap \Sigma_{(a)}$ is given by the point $\vec k=0$, whereas $\Sigma_>\cap \Sigma_{(a)}$ corresponds
to
\begin{eqnarray}
\vec k=\bar \omega \vec v. \label{vec v}
\end{eqnarray}
To verify these, we first note that $k^2=0$ in $\Sigma_{(a)}$ implies $\omega=0$ that means $\pmb v \cdot \pmb k=0$. Since $\pmb k$ is null and $\pmb v$ is timelike, we get $\pmb k=0$. Similarly, for $\Sigma_>\cap \Sigma_{(a)}$,
if we set $\omega=\omega_>$ in $\Sigma_{(a)}$, we get
\begin{eqnarray}
\frac {g^2 \chi \omega_0^2 \omega^2}{\omega^2-\omega_0^2}= \frac {\bar \omega^2}{4\pi},
\end{eqnarray}
so that the points in the intersection must satisfy
\begin{eqnarray}
k^2-\bar \omega^2=0,
\end{eqnarray}
where $k^0 v^0=\bar \omega+\vec k \cdot \vec v$.
In the frame where $\vec v=0$ this has solution $\vec k=0$, and boosting at a general $\vec v$ gives (\ref{vec v}). This can also be determined directly. Assuming $\vec v\neq \vec 0$, let $\hat v$ the direction of $\vec v$. We can
write
\begin{eqnarray}
\vec k=\alpha \hat v+\beta \hat v_\perp,
\end{eqnarray}
where $\hat v_\perp$ is a normalised vector perpendicular to $\hat v$. The equation for the intersection thus becomes
\begin{eqnarray}
\left( \frac {\bar\omega}{v^0}+\alpha \frac {|\vec v|}{v^0} \right)^2=\bar \omega^2+\alpha^2+\beta^2. 
\end{eqnarray}
After some manipulations it takes the equivalent form
\begin{eqnarray}
0=\beta^2+ \left( \bar\omega |\vec \nu|-\alpha \sqrt {1-\vec \nu^2} \right)^2,
\end{eqnarray}
which gives $\beta=0$ and (\ref{vec v}). The intersection points are schematically depicted in figure \ref{figura dispersion}. 

\subsection{The test functions}
In order to avoid the above singularities, and get well defined fields and propagators, we choose as test function space the set
\begin{eqnarray}
\mathcal S_{\tens v} (\mathbb R^4):=\{ f\in \mathcal S(\mathbb R^4) | g(\tens k)=(\omega^2-\bar\omega^2)^{-2} \mathcal F[f](\tens k) \Rightarrow g\in \mathcal S(\mathbb R^4) \},
\end{eqnarray}
where $\mathcal F$ is the Fourier transform. It is immediate to verify that with this choice all fields and propagators are well defined distributions. Moreover, the set of such spaces transform covariantly under the Lorentz transformation,
and each of them is dense in $L^2(\mathbb R^4,d\nu_a)$, see (\ref{measure}), as a consequence of the Wiener Tauberian theorems \cite{wiener}. \\
A comment is in order. Indeed, all singularities are generated by commutators of the kind (\ref{commutatore singolare}), each one giving a singularity of order one. The second order pole arises because the $a_0$ coefficients in $\tens A$ have already a first 
order singularity, and a second one is added by the commutator. The singularity in the field has been introduced by the shift (\ref{shift}), which has been introduced as a redefinition of the mode $a_3(\vec k)$ in order to get the commutators
$[a_3(\vec k), a_3^\dagger(\vec q)]=0$. So, we may wonder if such a singularity is simply due to a bad redefinition of the fields, then questioning if such redefinition was allowable.
However, it is easy to check that if we were not performing the shift (\ref{shift}), then we would not have a singular expression for the field, but we would get
\begin{eqnarray}
[a_3(\vec k), a_3^\dagger(\vec q)]=\left[ \xi +4\pi \frac {(\omega_+^2-\omega_0^2)(\omega_+^2-\bar\omega^2)-8\pi \omega_+^2 \omega_0^2 g^2\chi}{(\omega_+^2-\bar\omega^2)^2} \right] \frac { |\vec k|}{\omega_+^2} \delta^3(\vec k-\vec q),
\end{eqnarray}
in place of zero. This shows that the second order pole appears in the propagator independently from the shift (\ref{shift}).

\section{Conclusions and outlooks}\label{sec: conlusions}
We have provided a full relativistic quantum theory of the electromagnetic field interacting with an homogeneous and isotropic medium. The fields, which we now indicate generically with $\tens \Phi$, are represented as operator valued distributions 
\begin{eqnarray}
\mathcal S'_{\tens v} (\mathbb R^4)\otimes {\mathcal L}(\mathcal F_{\tens v}),
\end{eqnarray}
where ${\mathcal L}(\mathcal F_{\tens v})$ is the set of linear operators on the separable Hilbert space $\mathcal F_{\tens v}$. Indeed, one gets a family of representations parameterised by
$\tens v\in\mathcal B=SO(1,3)^{\uparrow}/SO(3)\simeq V^+=\{\tens x \in \mathbb R^{1,3}|\ \tens x \cdot \tens x >0 \}$. In this sense the vacuum state is not the unique state invariant by translations, but we have a family
of vacuum states which are however substantially identified by unitary maps. Indeed, on each $\mathcal F_{\tens v}$ it is not represented the whole Poincar\'e group but only the little group of $\tens v$, the set of Poincar\'e transformations
which leave $\tens v$ invariant. The remaining Poincar\'e transformations $\tens v\mapsto \tens \Lambda \tens v$, the boosts relating different inertial frames, essentially realise the unitary equivalences 
$U(\tens \Lambda): \mathcal F_{\tens v} \rightarrow \mathcal F_{\tens \Lambda \tens v}$. Each vacuum state $\Omega_{\tens v}=|0\rangle_{\tens v}$ is unique in $\mathcal F_{\tens v}$ and cyclic for it. We have realised the theory
by means of the correspondence principle, so starting from the equal time canonical commutation relations. This, combined with Lorentz covariance, ensures that microcausality (or locality) is satisfied:
$[\tens \Phi(f), \tens \Phi(g)]=0$ for any $f,g\in \mathcal S_{\tens v} (\mathbb R^4)$ whose supports are spatially separated.\\
It may be interesting to compare these characteristics with the standard Wightman axioms, which we report here for convenience, \cite{Strocchi}:
\begin{enumerate}
\item The states are described by vectors of a separable Hilbert space $\mathcal H$; 
\item Energy momentum spectral condition:
\begin{itemize}
\item[a.] The space-time translations are a symmetry of the theory, and consequently they are described by (strongly continuous) unitary operators $U(a)$, $a\in \mathbb R^4$, in $\mathcal H$;
\item[b.] The spectrum of the generator $P_\mu$ is contained in the forward closed cone $\bar V_+$;
\item[c.] There is a vacuum state $\Omega$, with the property of being the unique translationally invariant state in $\mathcal H$;
\end{itemize}
\item The theory is formulated in terms of fields $\{\Phi \}$ which are operator valued tempered distributions (densely defined) in $\mathcal H$ (with $\Omega$ in the domain of the polynomials $P$ of the smeared fields $\Phi(f)$) and the
vacuum is a cyclic vector with respect to $P$, namely $\mathcal H=\overline{\{P\Omega\}}$;
\item Covariance:
\begin{itemize}
\item[a.] The Lorentz transformations are described by (strongly continuous) unitary operators $U(\tens \Lambda)$, $\tens \Lambda\in SO(1,3)_+^{\uparrow}$ (the restricted Lorentz group); 
\item[b.] The fields transforms covariantly under the Poincar\'e transformations $U(\tens a, \tens \Lambda)\equiv U(\tens a)U(\tens \Lambda)$,
$$
U(\tens a, \tens \Lambda)\Phi_J(\tens x)U(\tens a,\tens \Lambda)^{-1}=S_J(\tens \Lambda^{-1}) \Phi_J (\tens \Lambda \tens x+\tens a),
$$
with $S_J$ a finite dimensional representation of $SL(2,\mathbb C)$, the universal covering of $SO(1,3)_+^{\uparrow}$;
\end{itemize}
\item Microcausality or locality: The fields either commute or anticommute at spacelike separated points
$$
[\Phi_J(f),\Phi_K(g)]_\pm=0, \qquad \text{if $f,g$ have support in spacelike separated regions,}
$$
where the $\pm$ stands for anti- and commutator respectively.
\end{enumerate}
We see that while axioms 1 and 2a are satisfied, the first difference is in axiom 2b: in our model the spectrum is not contained in $\bar V^+=\{\tens x| x^0\geq 0, \tens x^2\geq 0\},$ since the oneparticle spectrum contains the branches $\Sigma_>$,
$\Sigma_2$. Since the branch $\Sigma_2$ is well-known in experimental physics, 
it must be included in the spectrum defining the fields. We shall turn on this point later.
Axiom 2c is weakened in our case, since we have not a unique invariant vacuum $\Omega$, but a family of (equivalent) translationally invariant vacua $\Omega_{\tens v}$, each one unique in $\mathcal F_{\vec v}$. But this is not a novelty 
peculiar of our model, since it must happen each time a preferred frame is selected, e.g. when an external field is present, as previously discussed. 
Axiom 3 is also slightly modified by restricting the set of test functions to the Schwartzian functions whose Fourier transform vanish on suitable singular loci. Since these loci have zero Lebesgue measure, it follow from the Wiener Tauberian theorems
that these spaces remain dense. It is worth noting that the singularities we have chosen to smear out involve only the unphysical gauge components $\tens d\sigma$ of the fields, and are thus expected not to contribute to physical quantities.\\
Axiom 4 again is respected in a weaker form: on each $\mathcal F_{\tens v}$ is represented only a subgroup $SO(3)_{\tens v}\subset SO(1,3)^{\uparrow}_+$, isomorphic to $SO(3)$, by strongly continuous unitary operators. The remaining transformations
extend these to an action of unitary operators over a family of Hilbert spaces.
Finally, microcausality is realised with the minus sign, since we are working with bosonic fields, thanks to the covariance and the implementation of the equal time canonical commutation relations. Microcausality may be not sufficient to grant causality,
if axiom 2b is not satisfied. Nevertheless, this is not our case, where causality is respected, but, again, it could not be a surprise, see \cite{watson}.\\
This way we reached the goal of realising the exact quantisation of a nontrivial physical system, in the Heisenberg picture. There are several possibilities for extending this work. First of all, the model can also be quantised by means of the 
path integral methods. Even though it is a less rigorous way from the mathematical point of view, in a sense it sheds some new light on the quantisation of the system, and deserves investigation. This topic is presented elsewhere \cite{Path-Hopfield}.
An interesting generalisation would be to improve the model from the point of view of the physical properties of the dielectric medium, which in our minimal realisation are essentially absent. However, at the moment this task is out of our
goals. A simpler step would be to include interactions, inhomogeneities and non linearities. For example, in order to improve our analysis of the analogue Hawking effect in dielectrics we need to modify the model enough in order to include
at least some nonlinear effects, as the Kerr effect. More in general, it would be interesting to investigate other representations, like, for example, the thermal ones. All these topics are under investigation.
Our model can be extended to the case of an arbitrary number of polarisation fields, in a quite straightforward way. Another interesting possibility would be to cure singularities at resonances by including absorption. This may become relevant if one 
needs to investigate regions not too far from the resonance, so that the absorption becomes important. Far from the resonances the relativistic Hopfield model is enough for describing light in a transparent medium.


\newpage
\begin{appendix}
\section{Deduction of (\ref{quarantadue})} 
We need to solve the equation
\begin{eqnarray}
\mathcal M \tilde {\tens A}^F=\tens 0,
\end{eqnarray}
where 
\begin{eqnarray}
\mathcal M= 
\left[ \frac {k^2}{4\pi} -\frac {g^2\chi \omega_0^2 \omega^2}{\omega^2-\omega_0^2} \right] \mathbb I_4 +\frac {g^2\chi \omega_0^2 }{\omega^2-\omega_0^2} 
[\omega \tens v \otimes \tens k- k^2 \tens v \otimes \tens v].
\end{eqnarray}
Since $\tens k \cdot \tens v=\omega$, $\tens k$ is in the kernel of the last parenthesis. Also, being $\tens e_i \cdot \tens k=\tens e_i \cdot \tens v=0$, $\tens e_i$ are also in the kernel of the same matrix. Thus, $\mathcal M$ acts on these 
terms only by means of the first part which is proportional to the identity. The coefficient of proportionality is thus an eigenvalue with multiplicity 3. The fourth eigenvalue can be computed by observing that the image of the second
matrix is proportional to $\tens v$. Since the first part is proportional to the identity, this means that $\tens v$ is the fourth eigenvector:
\begin{eqnarray}
\mathcal M \tens v=k^2 \left[ \frac 1{4\pi} -\frac {g^2\chi \omega_0^2 }{\omega^2-\omega_0^2}  \right] \tens v.
\end{eqnarray}
The determinant of the matrix is the product of the four eigenvalues:
\begin{eqnarray}
\det \mathcal M= \left[ \frac {k^2}{4\pi} -\frac {g^2\chi \omega_0^2 \omega^2}{\omega^2-\omega_0^2} \right]^3 \frac {k^2}{4\pi} \frac {\omega^2-\bar\omega^2}{\omega^2-\omega_0^2}.
\end{eqnarray}
The supports of the equation are thus defined by $\det \mathcal M=0$.\\
The component 
\begin{eqnarray}
\left[ \frac {k^2}{4\pi} -\frac {g^2\chi \omega_0^2 \omega^2}{\omega^2-\omega_0^2} \right]^3 =0
\end{eqnarray}
corresponds to 3 different eigenvectors, $\tens e_1, \tens e_2, \tens k$, so it defines the solution
\begin{eqnarray}
\tens f(\tens k)=\delta \left( \frac {k^2}{4\pi} -\frac {g^2\chi \omega_0^2 \omega^2}{\omega^2-\omega_0^2} \right) \left[ a_1 (\tens k) \tens e_1
+ a_2 (\tens k) \tens e_2 + a_1 (\tens k) \tens k \right].
\end{eqnarray}
The other support pieces, $k^2=0$ and $\omega^2-\bar\omega^2=0$, correspond to the eigenvector $\tens v$, so they define the solution
\begin{eqnarray}
\tens g(\tens k)=(\omega^2-\omega_0^2)a_0(\tens k) \delta (k^2) \tens v+\tilde b_3 (\tens k) \delta (\omega^2-\bar \omega^2) \tens v.
\end{eqnarray}
Putting all together and defining $\tilde b_3(\tens k)=: 4\pi g b_3(\tens k)$, we get (\ref{quarantadue}). 

\

{\bf Remark:} the reason for defining $b_3$ is to get the right behaviour in the limit $g=0$. 

\

{\bf Remark:} the same result can be obtained directly from (\ref{A}), by considering that
\begin{align}
\int\frac {d^4 k}{(2\pi^3)} \omega &\left[\omega^2-\omega_0^2- \frac \xi{4\pi} (\omega^2-\bar\omega^2)  \right] a_0(\tens k) \tens k \delta^{(1)} (k^2)e^{-i\tens k\cdot \tens x}\cr =&
\int\frac {d^4 k}{(2\pi^3)} \left[\omega^2-\omega_0^2- \frac \xi{4\pi} (\omega^2-\bar\omega^2) \right] a_0(\tens k) \frac 12 \tens k v^\mu \frac {\partial}{\partial k^\mu}\delta(k^2) e^{-i\tens k\cdot \tens x} \cr
=&\frac i2 \int {d^4 k}{(2\pi^3)} \left[\omega^2-\omega_0^2- \frac \xi{4\pi} (\omega^2-\bar\omega^2) \right] a_0(\tens k) (\tens v\cdot \tens x) \tens k \delta(k^2) e^{-i\tens k\cdot \tens x}\cr
&-\frac 12 \int {d^4 k}{(2\pi^3)} \left[\omega^2-\omega_0^2- \frac \xi{4\pi} (\omega^2-\bar\omega^2) \right] a_0(\tens k) \tens v \delta(k^2) e^{-i\tens k\cdot \tens x}+\ldots,
\end{align}
where we have integrated by parts, and the first line of the last expression comes out from the derivative of the phase, the second line from the derivative of the vector $\tens k$, and the ellipses, which are the derivative of the remaining
factor, stay for harmonic pure gauge terms. 
 

\end{appendix}

\newpage

\end{document}